\documentclass[12pt]{article}
\usepackage{natbib}
\usepackage{amsmath}
\usepackage{amssymb}
\usepackage{amsthm}
\usepackage{graphicx}
\usepackage{graphics}
\usepackage{latexsym}
\usepackage{eucal}
\usepackage{epsfig}
\usepackage{mathrsfs}
\usepackage{setspace}
\usepackage{tabularx}
\usepackage{rotating}
\usepackage[none]{hyphenat}

\title{Large-Scale Simulating of RNA Macroevolution by an Energy-Dependent Fitness Model}
\begin{document}
\maketitle

\newpage
Abstract -- Simulated nucleotide sequences are widely used in theoretical and empirical molecular evolution studies. Conventional simulators generally use fixed parameter time-homogeneous Markov model for sequence evolution. In this work, we use the folding free energy of the secondary structure of an RNA as a proxy for its phenotypic fitness, and simulate RNA macroevolution by a mutation-selection population genetics model. Because the two-step process is conditioned on an RNA and its mutant ensemble, we no longer have a global substitution matrix, nor do we explicitly assume any for this inhomogeneous stochastic process. After introducing the base model of RNA evolution, we outline the heuristic implementation algorithm and several model improvements. We then discuss the calibration of the model parameters and demonstrate that in phylogeny reconstruction with both the parsimony method and the likelihood method, the sequences generated by our simulator, rnasim,  have greater statistical complexity than those by two standard simulators, ROSE and Seq-Gen, and are close to empirical sequences.\\


\newpage
In recent years, simulations are playing increasing roles in our understanding of molecular evolution and in evaluating data analysis procedures such as phylogeny estimations, sequence alignments, and gene family diversification. Various Monte Carlo methods exist for simulating sequences \citep[page 302-305]{Yang2006cme}, and are implemented in many simulators such as the popular Seq-Gen \citep{Rambaut1997}, EVOLVER \citep{Yang1997} and ROSE \citep{Stoye1998}. These simulators were built upon the theory of continuous-time Markov chain. For nucleotide evolution, the state space for the chain is $S = \{A, T, G, C \}$ or $S = \{A, T, G, C, - \}$ if insertion and deletion are allowed. Each site evolves according to an instantaneous rate matrix $Q$. Given a site is in state $i$ at time $0$, its state at time $t$ is probabilistically determined by $Q$ and the transition-probability matrix $P(t) = e^{Qt}$. For time-reversible Markov chains, many different transition constraints have been modeled for Q, e.g.,
JC69 \citep{jukes1969epm}, Kimura80 \citep{kimura1980sme}, HKY85 \citep{hasegawa1985dha}, etc. Each site is assumed to mutate independently with possibly different rates that can be modeled by probability distributions such as the gamma distribution \citep{yang1993mle, yang1994mle}. Lineage-specific rates can also be introduced by the covarion-model and adds another layer of complexity \citep{galtier2001mlp, huelsenbeck2002tcm, guindon2004mss}.\\

The site-independence assumption, though facilitating simulation, is a simplification of biological reality and can impair assessment of phylogenetic accuracy \citep{huelsenbeck1999ens}. A site usually evolves under the constraints of other sites, either locally or remotely. The site-independence constraint has been relaxed in some studies. For example, higher order Markov chain was used to model local site dependence \citep[page 117-132]{tavare1989}, and autocorrelation for non-overlapping correlations among sites have also been introduced \citep{schoniger1994sme}. In particular, the secondary structure, especially the base-pairings in stem region, of RNA requires the consideration of site dependence and indeed many models have been developed \citep[e.g.,][]{renee1994mlm, rzhetsky1995esr, muse1995ead, hudelot2003rbp, smith2004ems, gesell2006sse, yu2006das}. Case studies indicate that incorporating secondary structure of RNA in the tree estimation procedure improves phylogeny accuracy \citep{telford2005crs, erpenbeck2007pau}.\\

Ideally, a more realistic simulation should take into account the genotype-phenotype-fitness relationships of sequences to organisms and their consequences on macroevolutionary dynamics. While a detailed genotype-phenotype-fitness model is difficult, recently RNA molecules have been used in evolutionary simulations
\citep[e.g.,][]{ancel2000pea, kupczok2006dsr}. For an RNA gene, its primary sequence is the genotype and secondary structure is one kind of phenotype. Natural selection leaves its footprints on genotypes by selection on phenotypic fitness. Because the phenotypic fitness of an RNA is determined by many factors including stability, geometry, chemistry and other \emph{in vivo}  molecules and environment, there is no obvious way to connect genotype to phenotypic fitness. However, one commonly used proxy of fitness is the folding free energy of RNA secondary structures. Under certain physiological conditions (temperature, ionic solution, etc.), the free energy is determined by the enthalpy and entropy of base pair stacking and loops, with further geometric constraints such as end loop size \citep{zuker1981ocf, zuker1989fas, hofacker1994ffa}. The site dependence arises naturally as a (complex) function of this thermodynamic property. Because the free energy change caused by a mutation is determined by nucleotides at the mutated site and many other interacting sites, each site has a rate matrix $Q$ that varies both temporally and spatially, adding complexity to simulation. These RNA related properties were exploited in some recent studies \citep{yu2006das, thorne2007pgw}. In this paper, we present a simulation-based model of the macroevolution of RNA incorporating a fitness-based selective dynamics for fixation of new mutations using thermodynamic free energy as a proxy for fitness. We first describe an approximation scheme for implementing an efficient selective dynamics, then calibrate simulation parameters based on empirical data. Finally we demonstrate that the simulated sequences have statistical characteristics for phylogeny estimation that more closely resemble empirical data compared to other macroevolutionary simulators, suggesting its utility for testing phylogeny and molecular evolution estimators.\\

\newpage
\begin{center}
\begin{large}F\end{large}ITNESS-DEPENDENT \begin{large}R\end{large}NA \begin{large}E\end{large}VOLUTION \begin{large}S\end{large}IMULATION\\
\noindent\textit{Fixation Rate of a Mutation}\\
\end{center}

\begin{figure}
\begin{center}
\epsfig{file=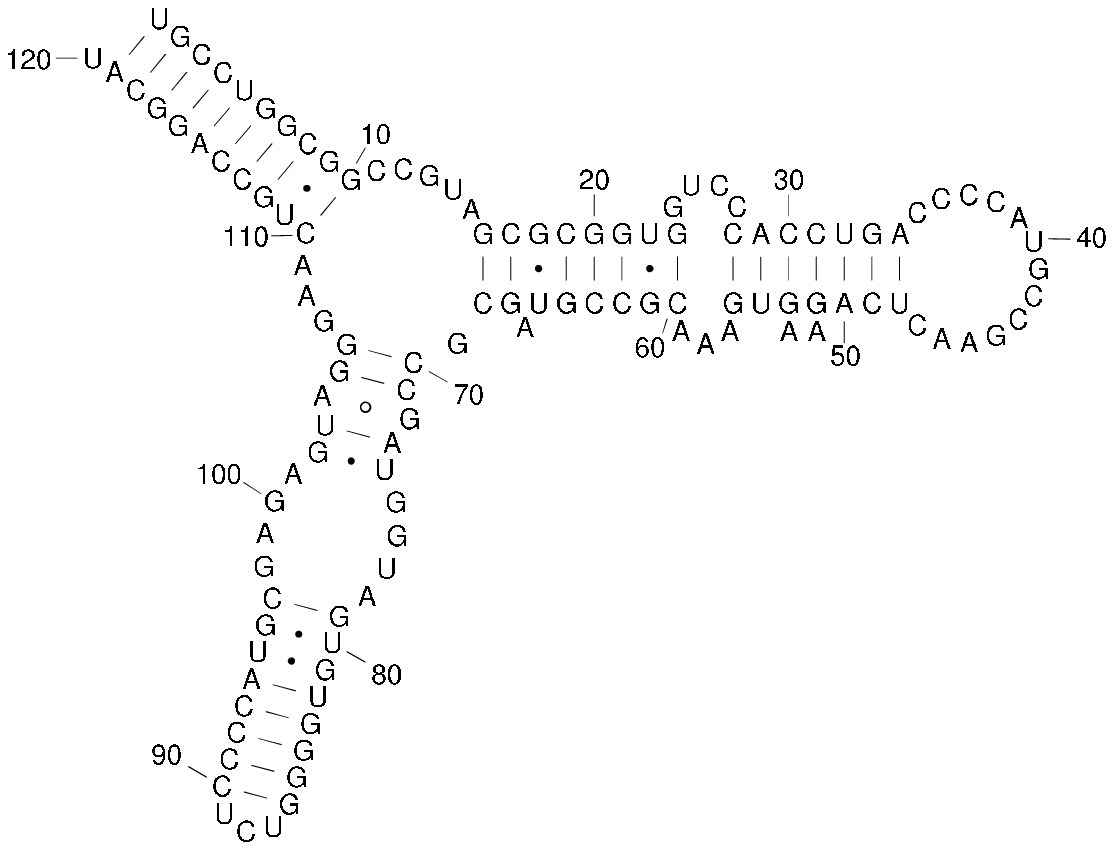,height=2.6in}
\begin{scriptsize}
\begin{verbatim}
>E.coli_5S_rRNA
UGCCUGGCGGCCGUAGCGCGGUGGUCCCACCUGACCCCAUGCCGAACUCAGAAGUGAAAC
GCCGUAGCGCCGAUGGUAGUGUGGGGUCUCCCCAUGCGAGAGUAGGGAACUGCCAGGCAU
((((((((((.....((((((((....(((((((.............))))..)))...)
))))).)).((.((....((((((((...))))))))....)).))...)))))))))).
\end{verbatim}
\end{scriptsize}
\end{center}
\caption{\label{Figure1}}
\end{figure}

\indent\indent In our simulation, we assume that the functional structure of an RNA is approximated by the secondary structure of the molecule, which can be represented by a convenient bracket format (Fig.~\ref{Figure1}). RNA's can be mutated by nucleotide substitution, insertion and deletion. If a mutation occurs in a hydrogen-bonded stem region, the favorable bonding energy may be reduced with either weakening of the stem configuration or a reduction of the stem length. Two consecutive substitutions may change one hydrogen-bonded pair of nucleotides into a different bonded pair in what is called a compensatory mutation. In a population, some mutations fix due to positive selection or random drift while some mutations will be lost due to negative selection or random drift. The probability for a mutation to fix is determined by many factors including the effective population size and the fitness changes it brings. In our model we first establish the fixation probability using what we call a pseudo-thermodynamic approach. In the following, the terms advantageous, neutral and deleterious will be used to denote fitness variants in relation to current fixed genotype.\\

Let $E$  be the folding free energy of an RNA molecule M. Let M mutate into M$'$ with free energy $E'$. The free energy change is

\begin{equation}
\Delta E = E' - E
\end{equation}

Here we first discuss a base model where a thermodynamically more stable molecule is considered fitter. Later, we discuss generalization to stabilizing selection around an optimal free energy value. Under the stability fitness model we have

\begin{displaymath}
\Delta E  \left\{ \begin{array}{ll}
<0 & \textrm{advantageous mutation}\\
=0 & \textrm{neutral mutation}\\
>0 & \textrm{deleterious mutation}
\end{array} \right.
\end{displaymath}

Assume that M has fitness 1, we model the fitness of $M'$ by

\begin{equation} \label{eq:t}
t = e^{-\alpha \Delta E}
\end{equation}

where $\alpha$ is a regularization factor that determines how strongly the free energy change affects fitness. The fitness change therefore is

\begin{equation} \label{eq:s}
s = t - 1
\end{equation}
For neutral mutations, such as mutating a nucleotide in the middle of a hairpin loop, $s=0$, i.e., there is no fitness change. For advantageous mutations, such as changing an unpaired A/G to a paired AU in stem regions,  $s>0$; for deleterious mutations, such as decoupling a base pair in a stem region, fitness decreases and $s<0$. The extreme of  a deleterious mutation is a lethal mutation for which $s=-1$. In our simulation, a hairpin loop of size 2 is prohibited in RNA secondary structure, any mutation causing that is considered lethal.\\

Assume we have a randomly mating diploid population of size $N$ with mutation rate $\mu$ (per allele per generation) and there is no dominance, that is, the relative fitness for the homozygous wildtype, the heterozygous mutant and the homozygous mutant is $1, 1+s$ and $1+2s$. Then from \citep{kimura1962} the fixation probability $p$ for a non-neutral mutation with fitness change $s$ ($s \ne 0$) is
\begin{equation} \label{eq:p_1}
p  = \frac{1-e^{-2  s}}{1-e^{-4 N s}}
\end{equation}
while if $s=0$, that is, if the mutation is neutral, fixation probability is $\frac{1}{2N}$. Averaging over the population the fixation rate is
\begin{equation}\label{eq:r}
r=2N\mu p
\end{equation}

\begin{center}
\noindent\textit{Basic Model of RNA Evolution}
\end{center}

\begin{figure}
\begin{center}
\epsfig{file=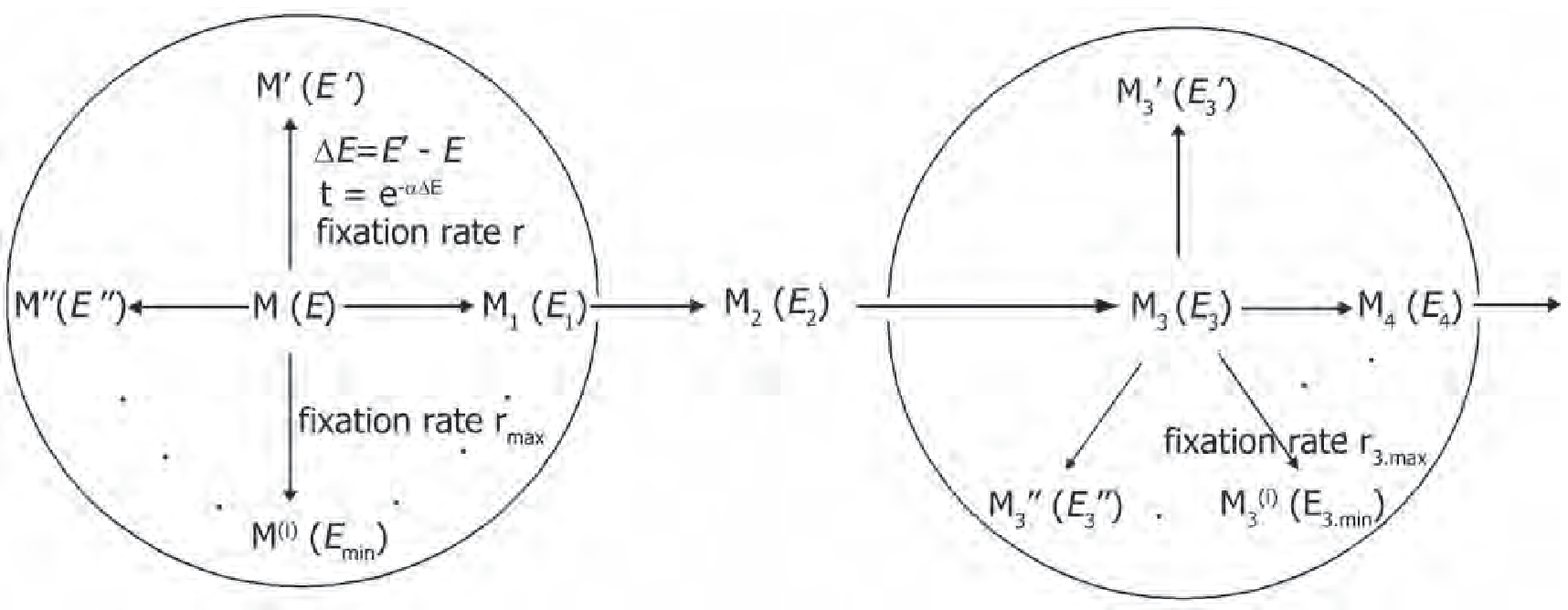, width=6.5in, angle=0}
\end{center}
\caption{\label{Figure2}}
\end{figure}

\indent\indent Our model of evolution is simulated over a rooted tree $\mathscr{T}$. A hypothetical ancestral RNA is used as the founder molecule at the root. For any edge of $\mathscr{T}$, the parental RNA mutates into a child RNA via fixation events along the edge. We define the \emph{mutant ensemble} $\mathscr{E}$ of an RNA molecule M to be the set of all its mutants that can mutate from M by a one-step mutation that may be a substitution, an insertion, or a deletion (Fig.~\ref{Figure2}). \\

Because mutants may differ in folding free energy and fitness, the ensemble is comprised of a collection of mutants with different fixation rates.  Let $r $ denote an individual fixation rate, then $R=\sum_{\mathscr{E}}{r}$ is the total fixation rate for an RNA. The mutation process is simulated by a time inhomogeneous stochastic process with state-dependent exponential waiting time with mean $1/R$. \\

For each fixation event, we need to appropriately draw from the ensemble $\mathscr{E}$. For an RNA molecule M, let $E_{min}$ be the lowest folding free energy of all mutants in $\mathscr{E}$. Mutants with $E_{min}$ have the highest fixation rate $r_{max}$. The relative fixation rate of a new mutant with fixation rate $r$ (computed from its free energy), compared to the fittest mutant, is $\gamma = r/r_{max} \le 1$. The relative fixation rate $\gamma$ can be used to implement a rejection-acceptance method for drawing from a probability distribution \citep[page 113]{devroye1986nur}. For each edge with length $e$ in units of generation of time in $\mathscr{T}$, we start with an ancestral RNA molecule, set $T=0$ and

\begin{enumerate}
\item Construct the mutant ensemble $\mathscr{E}$ for the current RNA molecule M.
\item Compute folding free energy for every member in $\mathscr{E}$.
\item Compute fixation rate $r$ for each member in $\mathscr{E}$ using corresponding free energy change. Record the maximal fixation rate $r_{max}$.
\item Calculate $R = \sum{r}$.
\item Draw a waiting time $t$ from the exponential distribution with mean $1/R$.
\item Add $t$ to time $T$ ($T = T+t$) which starts from the parental molecule. If $T > e$, stop simulation for this edge and set M as the child RNA molecule, otherwise, go to next step.
\item Randomly sample a mutant from $\mathscr{E}$ and accept it with probability $\gamma = r/r_{max}$. Continue until one mutant is accepted and replace M with the accepted mutant, then start again from step 1.
\end{enumerate}

\begin{center}
\noindent\textit{Heuristic Implementation}
\end{center}

\indent\indent  Direct implementation of the above simulation procedure for large-scale trees is infeasible because of several computationally demanding steps. In this section we replace these steps with heuristic implementations.\\

The first computational load comes from obtaining the secondary structures of mutants in $\mathscr{E}$ for an RNA molecule M. The popular program \texttt{RNAfold} from the Vienna RNA package  \citep[version 1.6.4]{hofacker1994ffa} folds up to moderately large RNA  sequences. However, the accuracy of folding drops considerably for large sequences. For example, it folds \emph{E.coli} 16S rRNA with 1542 nucleotides into rod-like structures with significantly more base pairs than actual secondary structure, apparently being too aggressive in forming stems (result not shown). In addition, it takes \texttt{RNAfold} five seconds to fold such a sequence on a 2.1 GHz Intel Pentium machine. If we restrict the indel length to be 1, for an RNA sequence of $n$ nucleotides, the size of $\mathscr{E}$ is about $8n$ ($3n$ mutants for substitution, $n$ for deletion, $4n$ for insertion) which would take \texttt{RNAfold} about 17 hours to compute the mutant ensemble of a 16S rRNA molecule, daunting even for a small-scale simulation. However, a mutant in $\mathscr{E}$ for RNA molecule M differs by only one nucleotide or a few nucleotides (depending on indel size) from M. The change in secondary structure is expected to be overwhelmingly local to mutation site(s). Therefore, instead of folding the whole sequence, we can locally update the secondary structure of a mutant using direct edits without carrying out thermodynamic minimizations. The folding free energy of the edited structure can be efficiently evaluated, by \texttt{RNAeval} in the Vienna RNA package,  to yield the fitness value. The possible edit operations that affect stem-loop structure are: coupling or decoupling a base pair at the end of a stem after a base substitution, decoupling a base pair after deletion of one base of the pair, base pairing a nucleotide at a bulge after inserting a base on its opposite side.\\

The second hurdle is to compute the waiting time $r$. In the basic strategy, we obtain $R$ by summing  $r$ over $\mathscr{E}$. In our model, fixation rate $r$ is a function of $N$, $\mu$, $\alpha$ and $\Delta E$. The first three are parameters held constant in a simulation, hence $r$ solely depends on $\Delta E$. It takes \texttt{RNAeval} in the Vienna RNA package about one minute to compute folding free energy $E$ for every mutant of the \emph{E.coli} 16S rRNA  using the direct-edit strategy. The running time is acceptable for a small-scale simulation but formidable for even a moderate-scale simulation. Here we propose to use a coarse-grained distribution of $\Delta E$ to save computation time. In this alternative, we first construct three distributions of $\Delta E$ for the three types of mutations: substitution, insertion and deletion. This is done by computing and storing $\Delta E$ of a set of reference mutant ensembles derived from $k$ initial reference RNA molecules in the simulation. We denote the reference distributions of free energy changes for substitution, insertion, and deletion mutations as $D_s$, $D_i$, and $D_d$, respectively. This approximation improves if the RNA set used for constructing distributions is larger and RNA molecules in the reference set are more diverse. The simple way to generate the reference set is to use the first $k$ RNA's in a simulation run and approximate the remainder using the ensembles enumerated for the first $k$ RNA. In our energy-directed simulation scheme, depth-first tree traversal for simulated data is favored over breadth-first traversal to give more diverse reference distributions. Afterwards, in computing $R$ for an RNA molecule M, we enumerate M's mutant types and compute individual $r$ using $\Delta E$ sampled from the corresponding distribution. Because the ensemble of M is  fairly large, errors of $r$ for individual mutants are averaged out and $\widehat{R}$, the estimate of $R$, is close to the actual value of $R$. In testing simulations on a 5000-taxa ultrametric tree, we used k=16 RNA mutant ensembles to construct the reference distributions and found that $\widehat{R}$  consistently differs from $R$ by less than 10\% even for RNAs on the terminal branches on which an RNA already differs from the root RNA by more than 100 fixed mutations. \\

Since we no longer compute folding free energy $E$ for each member in a mutant ensemble $\mathscr{E}$, we immediately encounter the problem of locating $E_{min}$ used to compute $r_{max}$ which in turn serves as the  normalization factor to compute the relative fixation rate $\gamma$ for a mutant in the ensemble $\mathscr{E}$. The value of $E_{min}$ is used to derive an efficient conditional sampling of the mutants (i.e., it is the dominating probability of the acceptance-rejection procedure). Here we approximate $E_{min}$ by using the minimum free energy value from the preset reference distributions with updates to account for distribution shift with new fixed mutations. For an RNA with folding energy $E$, we take the smallest value from the pool of stored $\Delta E$ for all three, $D_s$, $D_i$, and $D_d$, distributions and denote it as $min\{\Delta E\}$, which is always smaller than 0, and let
\begin{equation*}
E_{min}^{'} = E + min\{\Delta E\}
\end{equation*}
replace the role of $E_{min}$ for this RNA molecule. The approximation of $E_{min}$ by $E_{min}^{'}$ is not exact and two cases of problems could arise:

Case 1: $E_{min}^{'} \le E_{min}$ and $r_{max}^{'} \ge r_{max}$. In this case, the relative fixation rate $\gamma$ of any mutant with fixation rate $r_i$ decreases from $r_i/r_{max}$ to $r_i/r_{max}^{'}$ by a common factor $r_{max}^{'}/r_{max}$. In step 7, the sampling method is still valid since $\gamma \le 1$. The relative acceptance probabilities for all mutants stay unchanged, but the efficiency of the sampling is reduced.

Case 2: $E_{min}^{'} > E_{min}$ and $r_{max}^{'} < r_{max}$. In this case, although the relative acceptance probabilities are still the same for all mutants, for those with folding free energy smaller than $E_{min}^{'}$, $\gamma > 1$. This violates probability law. \\

\begin{figure}
\begin{center}
\epsfig{file=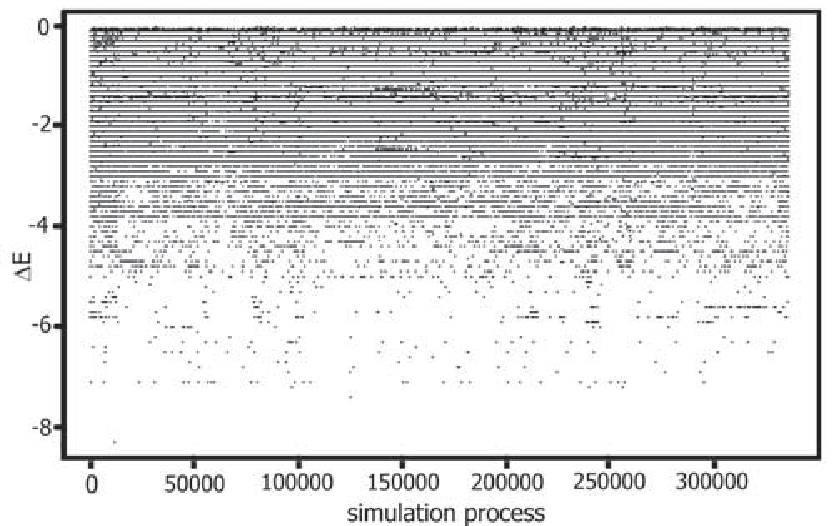, width=1.0\linewidth, angle=0}
\end{center}
\caption{\label{Figure3}}
\end{figure}

The easiest way to remedy the problem in the second case is to make $E_{min}^{'}$ smaller by subtracting some fixed factor $\epsilon$. If $\epsilon$ is large enough then $\gamma \le 1$ for all samples, while some moderate value will assure that $\gamma \le$ 1 for all but a small number of possible samples. The small number of samples with $\gamma > 1$ can be accepted with probability 1 without greatly distorting the sampling distribution. In practice, this folding energy underflow is a trivial problem if the set of RNA's used to construct the reference distributions of $\Delta E$ is sufficiently large, because mutants with large negative $\Delta E$ are rare (Fig.~\ref{Figure3}). In addition,  although decreasing $E_{min}^{'}$ causes more rejected samplings, the cost tends to be minimal. \\

To summarize, the heuristic algorithm is as follows. Starting from the root of a tree and on any edge with length $e$ for an RNA molecule M, set $T=0$,  $min\{\Delta E\}=0$ and the value of $k$, and

\begin{enumerate}
\item If the number of simulated RNA's is less than or equal to  $k$,
\begin{enumerate}
\item construct the mutant ensemble $\mathscr{E}$ of M;
\item compute folding free energy for every member in $\mathscr{E}$ and store the free energy difference $\Delta E$ between the mutants and M by mutation types (substitution, insertion and deletion) to create $D_s$, $D_i$, and $D_d$.
\item compare the minimum in $D_s$, $D_i$, and $D_d$ with  $min\{\Delta E\}$ and update $min\{\Delta E\}$ if necessary;
\item compute fixation rate $r$ for each member in $\mathscr{E}$ using corresponding $\Delta E$;
\end{enumerate}

Else (i.e., the number of simulated RNA's is greater than $k$)
\begin{enumerate}
\item count the number of possible types of mutations (substitution, insertion and deletion) for the given molecule.
\item sample $\Delta E$ from the stored reference distributions for different types of mutations, proportional to the possible types of mutations.
\item use sampled $\Delta E$ to compute fixation rate $r$.
\end{enumerate}

\item Calculate $R = \sum{r}$.

\item Draw a waiting time $t$ from an exponential distribution with mean $1/R$.

\item Add $t$ to time $T$ ($T = T+t$) which starts from the parental molecule. If $T > e$, stop simulation for this edge and set M as the child RNA molecule. Else, go to next step.

\item Randomly sample a mutant from $\mathscr{E}$, compute its folding free energy change $\Delta E$. Set $min\{\Delta E\} = \Delta E$ if $\Delta E < min\{\Delta E\}$. Estimate the maximal fixation rate $r_{max}$ using $min\{\Delta E\}$. Then Compute the fixation rate $r$ and relative fixation probability $\gamma = r/r_{max}$. Accept this mutant with probability $\gamma$. Continue the above procedures in this step until one mutant is accepted and replace M with the accepted mutant. Then start again from step 1.
\end{enumerate}

\begin{center}
\noindent\textit{Model Improvements}
\end{center}

\indent\indent In our model, evolution progresses as a simple two-step process. A mutation occurs to an individual and with a certain probability it eventually fixes or is lost from the population. In our base model, an insertion, a deletion and a substitution all have the same probability of occurring in the first step with the same base mutation rate $\mu$. We can introduce a variation by assuming that there is a constant ratio $\kappa$ between substitution and indel mutations. In practice, we found that if $\kappa \approx 20$, simulated sequences have statistically similar indel location distribution as aligned empirical sequences. The indel/substitution mutation model can be implemented by partitioning the mutation ensemble $\mathscr{E}$ into two parts and sampling as conditional distribution in relation to each part.\\

\begin{figure}
\begin{center}
\epsfig{file=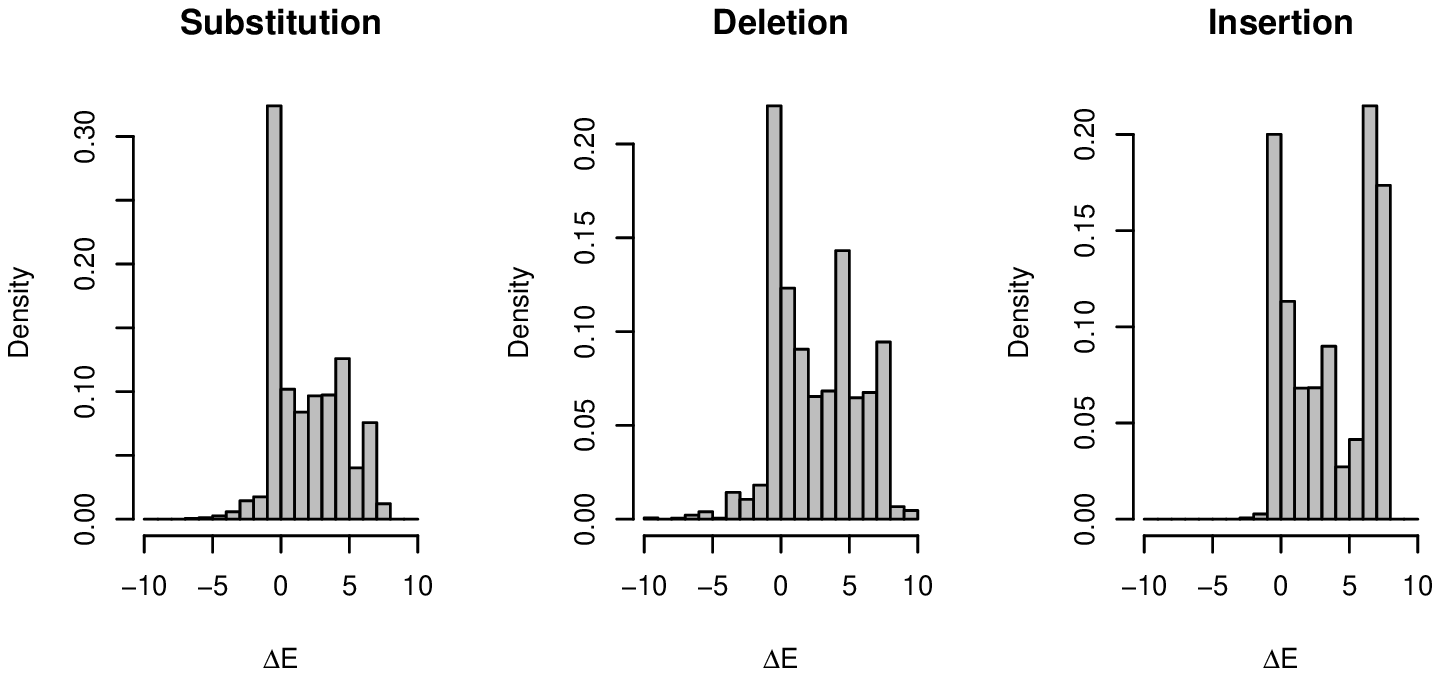,height=6in, angle=-90}
\end{center}
\caption{\label{Figure4}}
\end{figure}

For our base model, we either explicitly or implicitly calculate the mutation ensemble of a given RNA molecule. This necessitates that the indel size must be small to prevent combinatorial explosion of the size of $\mathscr{E}$. For a length $n$ RNA sequence, the size of its mutant ensemble $\mathscr{E}$ is $8n$ if the indel length is 1. If we allow insertion or deletion of up to $l (>1)$ nucleotides at one time, the number of insertion mutants alone explodes to $\sum_{i=1}^{l}{4^in}$, a large number even for small $l$.  For $n$ in the range of several hundred to a few thousand, we can no longer check every indel mutant to search for $min\{\Delta E\}$. Fortunately, the minimum $\Delta E$ of substitution mutations is smaller than $\Delta E$ for nearly all indel mutations. For example, for \emph{E. coli} 16S rRNA, the minimum $\Delta E$ of substitution mutations is smaller than $\Delta E$  for 99.99\% indel mutations (Fig.~\ref{Figure4}). So we can simply use the minimum $\Delta E$ of the substitution mutations to approximate $min\{\Delta E\}$ for all mutations. This computational simplification allows us to introduce more complex models of indel mutations. To generalize the indel model we incorporated a power-law distribution of the length of indels \citep{benner1993eas}, where the frequency of gaps with length $x$ is proportional to  $x^{-1.7}$.\\

In our model, the probability of mutation per site varies due to fitness-dependent fixation probabilities. However, we may also assume that input mutation rate may differ for each position. Here we introduce a weight variate $w$ modeled by a one-parameter gamma distribution with the probability density function
\begin{equation}\label{eq:gamma}
f(w|g) = \frac{ g^{g}}{\Gamma(g)} e ^{-g w}  w ^{g-1}  \qquad g >0, \quad w>0
\end{equation}
where $w$ has unit mean and variance $1/g$ \citep{yang1993mle}. To prevent extreme weight, we upper bound $w$ by a moderately small constant $c$. For a non-root RNA, a site  inherits $w$ from its homologous site in the immediate parental RNA. An inserted nucleotide is assigned a new weight that can be the weight of a neighboring nucleotide or the average of its two neighboring nucleotides. Weight $w$ of a site can be interpreted as the relative likelihood of this site to have a mutation. Therefore, a mutation previously with fixation rate $r$ now has fixation rate $wr$. Therefore, in the  simulation algorithm, we replace $r$ with $wr$, then compute total fixation rate $R_1 = \sum{wr} $ for substitutions and use $\kappa$ to get $R$. The sampling method changes accordingly. We first sample, using $\kappa$, the type of mutation. Let $\sum{w}$ be the total weight of sites that can have the chosen type of mutation, a site is then sampled with probability $w/\sum{w}$. A mutation on this site is accepted with probability $\gamma =\frac{r}{r_{max}c} $. Note that the constant $c$ is placed at the denominator to guarantee $\gamma  \le 1$. \\

The final extension concerns folding free energy. In the basic model, a mutant with lower free energy than the wildtype is considered advantageous. Given a large population and sufficiently long time, directional selection will result in highly stabilized molecules with very low free energy values. Such molecules show an over-zealous formation of base pairs such that the secondary structures are dominantly occupied by stems with few and small loops (data not shown). Along with long stems, the GC percentage of the molecule also increase to extreme values due to directional selection for stability. All of these features are rarely seen in function natural RNA. To avoid such unnatural continued directional selection, we introduced the idea of an optimal free energy $E_{opt}$. A global $E_{opt}$ is given at the beginning of a simulation. For an RNA with folding free energy $E$, if $E \ge E_{opt}$, then any mutant with lower folding energy is advantageous; if $E < E_{opt}$, such a mutant is deleterious. Therefore for a lineage starting from the root RNA, the descendent RNA molecule's folding energy first experiences directional selection towards $E_{opt}$. After this  ``burn-in'' process, it experiences stabilizing selection around $E_{opt}$.\\

\begin{center}
\begin{Large}R\end{Large}ESULTS\\
\noindent\textit{Parameter Calibration}
\end{center}

\indent\indent Our model uses a pseudo-thermodynamic model to connect the fitness effect $s$ with folding free energy change $\Delta E$ between a mutant and its wildtype, with the parameter $\alpha$ that regulates how strongly $\Delta E$ affects $s$ (Equation~\ref{eq:t}). We observe that for any RNA in a mutant ensemble $\mathscr{E}$, $\Delta E$ have values generally in the range of -10 to 10 (\emph{J/mol}). In addition, for all three mutation types, most mutants have folding free energy larger than or similar to the wildtype (e.g., Fig.~\ref{Figure4}). If lower free energy is favored, then most mutations are deleterious, neutral or nearly neutral, while advantageous mutations with large $| \Delta E |$ are extremely rare. Because any member in $\mathscr{E}$ differs from the wildtype by only one nucleotide or a stretch of several indels, we do not expect to see large fitness change $s$. Therefore, $\alpha$ must be small and
\begin{equation}\label{eq:s_approx}
 s = e^{-\alpha \Delta E} -1 \approx  -\alpha \Delta E
\end{equation}
Assuming observed $|\Delta E| \le 10$, the range of $s$ is approximately $[-10\alpha, 10\alpha]$ with small $\alpha$. \\

\begin{figure}
\begin{center}
\epsfig{file=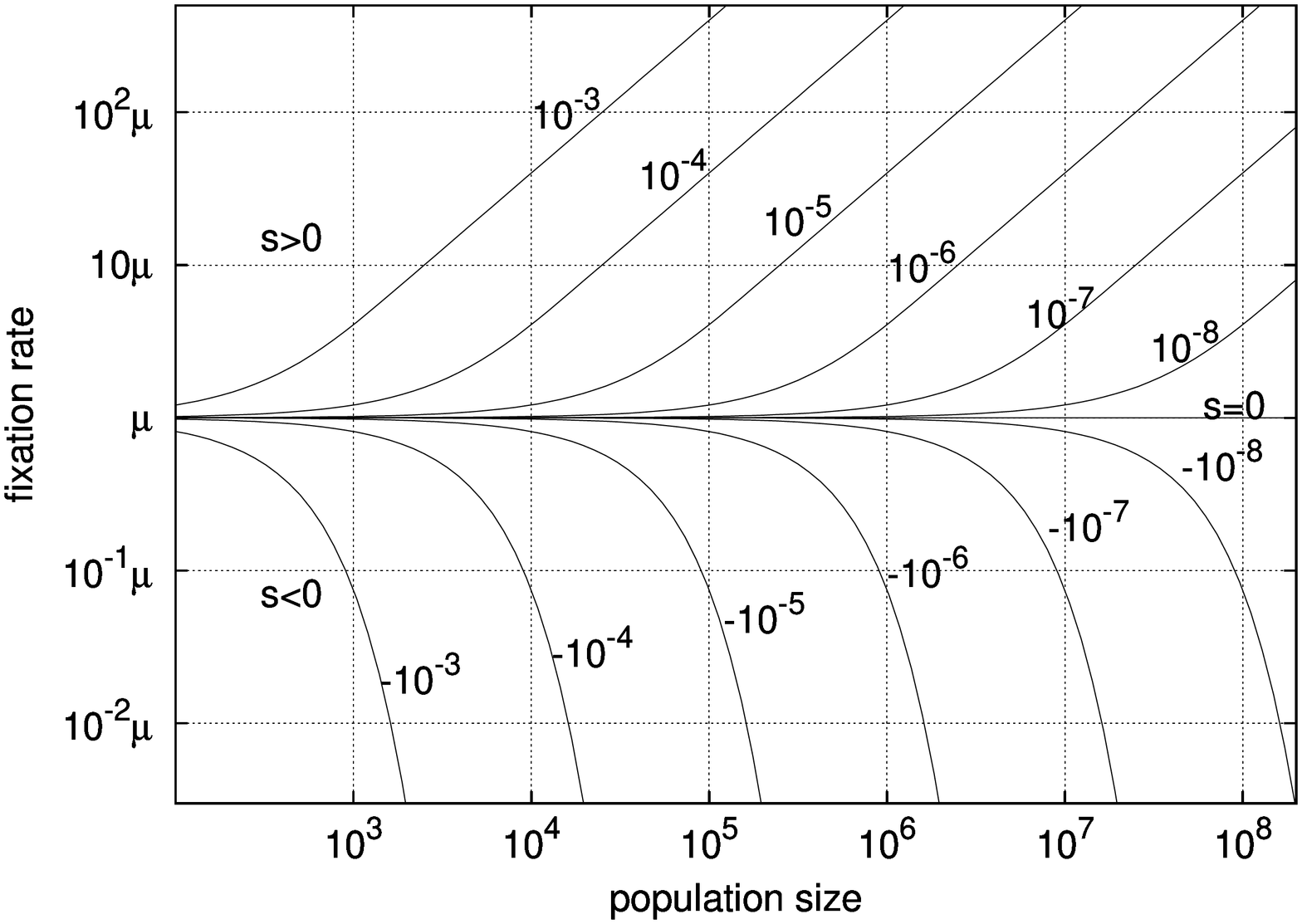,height=5in, angle=-90}
\end{center}
\caption{\label{Figure5}}
\end{figure}

The fixation rate $r$ is a function of $N$, $s$ and $\mu$, and is linear in $\mu$ (Equation~\ref{eq:r}). Its dependence on $N$ and $s$ is graphed in (Fig.~\ref{Figure5}), a well known result that we reproduced here for easy reference~\citep[page 187]{crow1986bcp}. We recap two phenomena. (a). An advantageous mutation, with $s>0$, has a larger fixation rate and is easier to fix in a larger population; vice-versa for deleterious mutation, as manifested in (Fig. 6). (b). For a given population size $10^n$, if the fitness change $s$ is on the order of $10^{-n}$, the fixation rate for an advantageous mutation is about 10- to 100-fold of the rate for a deleterious mutation. These observations, together with Equation~(\ref{eq:s_approx}) and the value range of $\Delta E$, can be used as guidelines in setting the two simulation parameters $N$ and $\alpha$. \\

There are four critical parameters in our model : $N$, $\alpha$, $\mu$ and $\kappa$. We have discussed the factors involved in setting of the first two parameters. Our initial calibration studies show that $\alpha = 10^{-4}$ and $N=10^4 \sim 10^5$ can be good default values to generate sequences that have similar statistical features in tree estimation to empirical sequences (data not shown). The parameter $\mu$ is a scaling parameter for all branches and under the exponential waiting time model, it is coupled to the branch length and thus the branch lengths can be seen to have units of $\mu t$, and we simply use a branch scaling factor in the simulator. The only other free parameter is $\kappa$, which can be set by the user. A smaller $\kappa$ introduces more indels. The default value of $\kappa$ is 20.

\begin{center}
\noindent\textit{Simulated Sequence Analysis}
\end{center}
\indent\indent In this part, we compare sequences simulated with our model with empirical sequences and those generated by two popular programs Seq-Gen and ROSE for their statistical behaviors with respect to phylogeny estimation algorithms. We first compiled a gapped empirical benchmark dataset of 1000 aligned small subunit ribosomal RNA sequences. A phylogeny was reconstructed from the gapped sequences and used as the guide tree for simulation. Both ROSE and rnasim have algorithms to generate ``true'' sequence alignment in which homologous sites among sequences are traced and automatically aligned, which made it easy to create gapped datasets. The ungapped datasets were created by removing gapped columns in alignments. Because Seq-Gen did not implement insertion and deletion, we only prepared ungapped datasets for it. All the gapped simulated sequences are of comparable sequence divergence level as the empirical dataset, so are the ungapped ones.\\

Both Seq-Gen and ROSE perform substitution according to rate matrices. In our comparison, Seq-Gen used the general time-reversible (GTR) model and ROSE used the HKY model since ROSE did not implement the GTR model. Rate variation among sites was enforced by a gamma distribution. The parameters for the rate matrices and the shape parameter for the gamma distribution were estimated from the empirical benchmark sequences. In addition, an \emph{E.coli} ssu rRNA was used as the root sequence for simulations. Our simulator, rnasim, used the same estimated gamma shape parameter. The other parameters were set to $N=5.0 \times 10^4$, $\alpha=10^{-4}$ and $\kappa=20$. The branch scaling factor was tuned to generate sequences at comparable level of divergence with the empirical benchmark sequences. Finally, both ROSE and rnasim used the same indel probability and indel length distribution. Details of data preparation and parameter calibrations are given in the Materials and Methods section.\\

The main goal of our simulator was to generate sequences with greater statistical complexity than standard simulators; ideally with similar complexity as empirical sequences. Measuring complexity in a relevant manner is a non-trivial problem. For example, any marginal statistical properties such as entropy of aligned positions, indel lengths, sequence composition, etc. can be ``tuned'' individually for all of the simulators to sufficiently emulate empirical data--in fact, as we did to calibrate the simulators. Neither Seq-Gen nor ROSE implements higher-order dependencies and therefore a comparison in this regard would not be useful. One of the key utilities of our simulator is as a test bed for phylogeny estimation algorithms, wherein previous simulation-based tests, as demonstrated later, were too simple. Therefore, here we examined the statistical nature of the sequences from the simulators as well as empirical data in establishing the complexity of the objective function landscape; that is, the complexity of the local peaks and valleys in relation to the tree score. Because of the computational costs of the investigations described below, we concentrated on the objective function landscape of the maximum parsimony function and maximum likelihood function. However, we hypothesize that the statistical nature of the objective function landscape will not differ greatly between different estimation methods.\\

We first compared the parsimony score convergence rate for gapped empirical benchmark sequences and the sequences generated by rnasim and ROSE (Seq-Gen was excluded in this test as it does not simulate gapped sequences). The simulated sequences were of the same length as the empirical data set and similar in sequence divergence. Heuristic tree searching was performed by PAUP* (version 4.0b10) with tree bisection-reconnection (TBR) branch-swapping procedure, during which all minimal trees were saved and used  as input to the branch swapping procedure. The search was monitored every second and the parsimony score of the current minimal tree was recorded. We expected that more complex datasets will induce more difficult search landscape and will take a longer time until a local optimum is found. We prepared three rnasim simulated datasets and three ROSE  datasets. In pilot experiments where we allowed the tree searches to proceed up to a week, the objective function always declined to a stable value within one hour. Therefore, along with the empirical benchmark dataset, each dataset was tree searched 10 times, with a time limit of 90 minutes in 1.5GHz Pentium machines with 512MB memory. We denote ``convergence time'' as the time after which the parsimony score no longer declines within the search time limit.  We found that the convergence time was not statistically different ($P$-value=0.42, unpaired $t$-test) between the empirical benchmark dataset (2099.40 $\pm$ 649.66 sec) and the rnasim datasets (1974.10 $\pm$ 319.27 sec). The ROSE datasets converged more quickly (1188.00 $\pm$ 359.72 sec), and was significantly faster than both the benchmark dataset  ($P$-value$=2.0\times10^{-4}$, one-sided unpaired $t$-test) and the rnasim datasets ($P$-value$<1.0\times10^{-4}$, one-sided unpaired $t$-test). We also examined the score convergence trajectories for the datasets and we noticed that the benchmark and rnasim trajectories are quantitatively similar: the parsimony score initially shows a rapid decrease, then settles into a very slow convergence, perhaps due to a plateau. In contrast, the ROSE trajectory decreases more gradually and over a broader time range without a clear sign of an optimality plateau. The time at which the parsimony score decreases by 10\%, 50\%, 95\%, and 99\% of the total score decrement, i.e., the difference between the starting score and the final score, is plotted in (Fig. ~\ref{Figure7}a). The ROSE datasets reach these four check points, especially the last three, much quicker than the benchmark dataset and the rnasim datasets. \\

We next tested ungapped sequences by removing the columns with gaps in the alignment. These sequences were shorter and had higher sequence identity than gapped sequences.  The sequences generated by Seq-Gen did not  have gaps and were also included. Again, we have one benchmark dataset and three datasets for each simulator. We performed 10 replicate estimates for each dataset using the same search settings as for the gapped sequences. The convergence time, in seconds,  for the empirical benchmark, rnasim, ROSE and Seq-Gen datasets were 3196.21$\pm 110.07$, $1037.30\pm 96.09$, $525.80\pm 51.48$ and $280.00\pm 9.04$, respectively. Three one-sided unpaired $t$-tests were done, and the $P$-values between benchmark and rnasim datasets, between rnasim and ROSE datasets, and between ROSE and Seq-Gen datasets were all smaller than $1.0\times10^{-4}$. The score convergence behavior (Fig.~\ref{Figure7}b) also demonstrates that the empirical benchmark dataset is the hardest for locating an optimal tree. In particular, it takes much longer for the parsimony score to reach 99\% of the total score decrement. The ROSE datasets and the Seq-Gen datasets exhibit much rapider descent while the rnasim datasets have a convergence rate in between the empirical dataset and the other simulated datasets.  The trajectory difference between the simulated datasets becomes more obvious at a higher sequence identity level suggesting that simulated datasets by ROSE and Seq-Gen become generally simpler with less divergence while simulated datasets by rnasim tend to maintain certain inherent level of complexity (Fig.~\ref{Figure7}c). \\

The parsimony score convergence rate measures how easy it is to find \emph{an} optimal tree. Because the number of binary trees with 1000 taxa is extremely large and an exhaustive checking of all trees is impossible, the resulting tree in a search may be a local optimal parsimony tree. Indeed, at least two optimal parsimony scores were found in the 10 experiments performed for each dataset above. In a geometry analogy,  we can view each possible tree as a point in a surface, its parsimony score for associated sequences being its altitude. The points are connected by branch swapping. A local optimum is a valley in this objective function landscape. We next measured the roughness of this landscape for different datasets, by examining the optimal parsimony scores in 1000 heuristic searches, done in PAUP*.  For the gapped comparison, the empirical benchmark dataset and one dataset for rnasim and ROSE were studied. For the ungapped comparison, an additional dataset for Seq-Gen was included. We used a constrained search procedure by storing only minimal tree at each step during the heuristic search. The procedure was repeated 1000 times for each dataset. We therefore obtained 1000 locally optimal parsimony scores and 1000 trees. The topological similarity between pairs of the 1000 trees was measured by Robinson-Foulds (RF) symmetric distance \citep{robinson1981}.\\

We summarize the results in Table 1. For the parsimony scores, we give the maximum, the minimum and the range of the scores. Each score was then subtracted by the corresponding minimum score to create a base adjusted score. The average and standard deviation (S.D.) of the base adjusted scores were given.
The maximum, the minimum, the range, the average and the S.D. of the 499,500 RF distances were also given without any base adjustment. For the gapped sequences, the rnasim dataset has an average base adjusted parsimony score of 17.9($\pm 9.4$), though smaller than 52.4($\pm 21.8$) of the empirical dataset, is much higher than 2.6($\pm 2.9$) of the ROSE dataset. Similarly, the average RF distance between trees obtained from the rnasim dataset is 130.6($\pm 21.0$), smaller than 369.2($\pm 36.9$) of the empirical dataset but significantly larger than 41.9($\pm 9.5$) of the ROSE dataset. To visualize the relative distance between the 1000 trees, we performed multidimensional scaling of the RF distance matrices of the gapped datasets \citep{gower1966sdp} and plotted the first two dimensions (Fig.~\ref{Figure8}a-d). We observe that the scatter pattern among trees for our simulated datasets resemble that of the benchmark datasets, while the ROSE dataset and the Seq-Gen dataset have much simpler patterns. Therefore, our simulated dataset has a rugged parsimony landscape that is somewhat smoother than the empirical benchmark dataset, but considerably rougher than the ROSE and the Seq-Gen datasets. \\

We then compared different datasets' behaviors in tree estimation using likelihood as the optimality criterion in PAUP*. We first compared different datasets' convergence rates. For each simulator, the same three gapped and three ungapped datasets used above, along with the gapped and ungapped empirical benchmark datasets, were tree searched 10 times each, but using only 100 sequences because of the considerably increased computational load of the likelihood calculations. The heuristic tree search also used tree bisection-reconnection (TBR) branch-swapping procedure and saved all minimal trees during search. Each search ran on 2.2GHz Pentium machines with 2GB memory and a time limit of 24 hours (nearly all searches finished earlier). For gapped sequences, we found that the convergence time between the empirical dataset ($21720\pm16573$ sec) and the rnasim simulated datasets ($21489\pm10118$ sec) is not statistically different (P-value=0.9581), while both are significantly longer (P-value$<10^{-4}$) than the ROSE  datasets ($4482\pm2660$ sec). For ungapped sequences, the convergence time is $11234\pm4508$ sec for the empirical dataset and $10025\pm3218$ sec for rnasim datasets, also statistically not different (P-value=0.3590). Both are significant longer (P-value$<10^{-4}$) than the ROSE simulated sequences ($1488\pm696$ sec) and the Seq-Gen simulated sequences ($1370\pm387$ sec). When comparing the  trajectories of convergence rate for gapped sequences (Fig.~\ref{Figure9}a) and ungapped sequences (Fig.~\ref{Figure9}b), we also see that the rnasim simulated dataset is more similar to empirical dataset than the ROSE  dataset and the Seq-Gen  dataset (ungapped sequences only). For the 10 tree searches running on a dataset, their likelihood scores have greater variation for the empirical dataset and the rnasim dataset than for the ROSE dataset and the Seq-Gen dataset.\\

We also measured the roughness of the likelihood objective function landscape (Table 2). The empirical dataset and one simulated dataset each for rnasim, ROSE (and Seq-Gen for ungapped sequences) were tree searched 1000 times. Each heuristic search ran for 90 minutes on a 3.0GHz Pentium machine with 6GB memory. Again we used the constrained search procedure by storing only minimal tree at each step during the search, then measured the likelihood scores of the 1000 resulting tree and the RF distance between all pairs of the 1000 trees. The likelihood scores were also base adjusted to compute their average and standard deviation. We found that for gapped datasets, the average likelihood score is $1453.06\pm 1392.00$ for the empirical benchmark dataset, and a close $1301.35\pm 1138.61$ for rnasim simulated datasets. Both are much larger than $14.26\pm 63.93$ for ROSE dataset. The RF distance between trees of empirical dataset ($75.49\pm18.88$) and between trees of rnasim dataset ($73.16\pm16.04$) are also close and are more than 10 times larger than that for the ROSE dataset ($6.22\pm10.48$). For the ungapped datasets, all 1000 tree searches returned the same tree for the ROSE dataset, and only a few very closely related trees (RF distance$\le$7) for the Seq-Gen dataset. The search returned more topologically different trees for the rnasim datasets and the empirical datasets than for the ROSE datasets and the Seq-Gen datasets, as further visualized by the multidimensional scaling for the ungapped datasets (Fig.~\ref{Figure8}e-h).

\begin{center}
\begin{Large}D\end{Large}ISCUSSION
\end{center}

\indent\indent We were motivated to develop the rnasim under the NSF funded CIPRES project (http://www.phylo.org/) where developments of large-scale computational infrastructure and novel algorithms called for realistically challenging benchmark datasets with known phylogenies. As noted by many others \citep{brudno2003} and from our own pilot studies, parametric simulators, i.e., simulators using a homogeneous stochastic process with a small number of parameters, tend to produce datasets that are more conducive to tree estimations as compared to empirical datasets. Such an effect maybe pronounced for more parametric estimators \citep{sanderson2000pp}. Our simulator attempts to generate more complex evolutionary dynamics by explicitly incorporating genotype-dependent fixation events through modeling of RNA secondary structures. Because the secondary structure depends on multiple sequence elements and because our fitness model depends on the free energy of the resulting secondary structure, different parts of the molecule experience differential fixation of mutations. Furthermore, because the secondary structure itself evolves through time, the rates of mutational fixation changes over time, creating a time inhomogeneous process. One further utility of our simulator is the generation of indels and the ability to keep track of the correct alignment, which can be used to test alignment algorithms.\\

We attempted to characterize the statistical properties of our simulated dataset by asking how the dataset relates to the generating tree graphs. In particular, we assessed how any dataset affects the computational difficulty of finding optimal trees. We conjectured that more complex datasets would generate an objective function landscape for which it is harder to find local optima and where the local optima might have more disparate tree topologies. Our investigations with empirical dataset suggest that indeed real datasets have such “rugged landscapes”. A previous unpublished study in our lab using the 228-taxon rbcL dataset ~\citep{hillis1996}  also suggested very rugged landscapes for empirical data. As shown above, standard parametric simulators tend to generate datasets that have much easier optimization properties. It was particularly striking that the maximum likelihood tree topology for 1000 separately estimated local optima were identical for the data generated by ROSE and nearly identical for the data generated by Seq-Gen (Table 2). This is in stark contrast to the diversity seen in the empirical dataset. The data generated from our simulator, while not quite reaching the complexity of empirical dataset, displayed closer emulation of the empirical objective function landscape. \\

Simulations remain one of key approaches to testing phylogeny algorithms and procedures for estimating molecular evolution parameters. It is clear that the complexity of the simulation model will have a great impact on the results of such tests. In particular, it is desirable to introduce complexity in a way that is not parameterized by smooth distribution families (e.g., gamma distribution), which on the surface seem to model complexity but in reality the dimensionality of the models are bounded by the number of parameters characterizing the distribution \citep{kim2000sho}. Our simulator attempts to overcome these problems using an explicit genotype-phenotype mapping approach. An extended version of our simulation using a 1,000,000 taxon-tree has been made available through the CIPRES project website, which we hope will provide a better benchmark for evaluating evolutionary estimation algorithms.

\begin{center}
\begin{large}M\end{large}ATERIALS AND \begin{large}M\end{large}ETHODS\\
\textit{Benchmark Dataset}
\end{center}

\indent\indent A collection of aligned nucleotide sequences of small-subunit rRNA were downloaded from European Ribosomal RNA database (http://www.psb.ugent.be/rRNA). To obtain a quality subset, a sequence was removed if it contained more than 15 non-AUGC nucleotides (unknown or ambiguous nucleotides) or it was visibly incomplete and had large gaps. In the alignment of retained sequences, non-AUGC nucleotides were replaced by gaps and any columns with more than 90\% gaps was removed, since otherwise the alignment was very long ($>6000$). In the end, we compiled an empirical benchmark dataset of 1,000 sequences with 73.0\% average, 46.7\% minimum and 99.9\% maximum pairwise identity. The alignment had 1933 columns with 598 ungapped columns. The three pairwise identities for the ungapped sequences were 87.9\%, 62.0\% and 100\%, respectively. \\

A phylogeny was reconstructed from the gapped benchmark dataset by PhyML~\citep[version 2.4.5]{guindon2003sb}. PhyML implemented a fast algorithm to estimate large phylogenies by maximum likelihood method. In our setting, the input tree was created by BIONJ and we used the general-time reversible (GTR) model for nucleotide substitution, with four substitution rate categories. The program estimated the proportion of invariable sites, the gamma shape parameter, the nucleotide frequencies and an instantaneous rate matrix. The phylogeny reconstruction was also done in HKY model with similar settings and the transition to transversion ratio was estimated.

\begin{center}
\textit{Simulated Datasets}
\end{center}

\indent\indent Using the GTR phylogeny of the benchmark dataset as the guided tree, we simulated sequences by Seq-Gen (version 1.3.2), ROSE (version 1.3) and rnasim. \emph{E.coli} small-subunit rRNA was used as the root molecule.

\begin{itemize}
 \item In rnasim, the gamma shape parameter was set to the estimated value 0.569 from the benchmark dataset. It admitted indels of size 1 to 6 whose frequencies were 0.5963, 0.1836, 0.0922, 0.0565, 0.0386 and 0.0328. This power law distribution was based on a study on protein \citep{benner1993eas} and was found to perform well in our simulation. The substitution and indel ratio $\kappa$ was set to 20 which equal probability for insertion and deletion. Other parameters were $\alpha=10^{-4}$, $N=5.0 \times 10^{4}$. $E_{opt}$ was set to be 10\% lower than the folding free energy of the root RNA molecule. Because rnasim keeps tracking of homologous sites, a ``true'' alignment of the 1,000 leaf sequences was obtained.
 \item In ROSE, we used the same gamma shape parameter 0.569, indel distribution, and the insertion, deletion and substitution proportion as in rnasim. Because ROSE did not support GTR model (it supports ``JC'',``HKY'',``F81'',``F84'',``K2P''), we used HKY model for the simulation. The transition and transversion ratio was set to the estimated value 2.72 from the benchmark dataset, so were the equilibrium nucleotide frequencies. ROSE also generated a ``true'' sequence alignment and we used it.
 \item In Seq-Gen, we also used the gamma shape parameter 0.569. Seq-Gen does not implement insertion and deletion. We used the GTR model for substitution with parameters (nucleotide frequencies, rate matrix) estimated from the benchmark dataset.
\end{itemize}

We first simulated gapped datasets using ROSE and rnasim. We only keep 1933 columns in the aligned sequences by randomly dropping excessive columns. By tweaking branch length, the simulated sequences had similar ($\pm1.0\%$) average and minimum pairwise identities to the empirical benchmark sequences. We then constructed ungapped datasets for all three simulators. The ungapped sequences were simulated directly by Seq-Gen. In ROSE and rnasim, we first simulated gapped sequences then randomly drew ungapped columns in the alignment to get the ungapped dataset. The order of columns was preserved. All ungapped datasets had 598 columns. Similarly, the ungapped sequences had similar ($\pm1.0\%$)  average and minimal pairwise identities to the ungapped benchmark dataset. Datasets with different sequence divergence levels from the benchmark dataset were also simulated and used to compare the three simulators.

\begin{center}
\textit{Sequence Evaluation}
\end{center}

\indent\indent A simulated or the empirical benchmark sequence dataset, either gapped or ungapped, was used in PAUP* (version 4.0b10) to reconstruct phylogenies with parsimony or likelihood being the objective function. Starting from 100 randomly generated trees, the heuristic algorithm was used to search for optimal trees with tree bisection-reconnection (TBR) being the branch swapping procedure. All minimal trees found during branch swapping were saved. All trees saved were used as input to the branch swapping procedure. All other settings use PAUP* default values. Search status was reported every second (CPU time) and the parsimony score of current minimal score tree was recorded. Experimental trials ran with time limit set to 7 days and detected little parsimony score decrease after 1 hour, so the running time was set to 90 minutes for all datasets. The evaluation trials ran on Pentium 1.5GHz machines with 512 MB memory. \\

A second procedure was applied to phylogeny reconstruction also with parsimony being the optimality criterion and TBR the branch swapping procedure. The search started from one randomly generated tree and saved only one minimal tree found during the branch swapping procedure. All other settings use PAUP* default values. No time limit was enforced. The procedure repeated 1,000 times for a dataset, generating 1,000 trees. Their parsimony scores were recorded. The pairwise Robinson-Foulds symmetric differences between all pairs of the 1000 trees were computed. For each dataset, the classical multidimensional scaling (MDS) procedure \citep{gower1966sdp} was applied to the RF distance matrix and the first two dimensions were saved to generate visualization of the RF distance between the 1000 trees.\\

The two procedures above were also used for simulated and empirical datasets, both gapped and ungapped,  with likelihood being the optimality criterion and three modifications. 1). Likelihood scores instead of parsimony scores were saved. 2). Because of dramatically increased computation load, only 100 sequences for each dataset were used and they were from the same subtree (of the guide tree) for all datasets. 3). For the first procedure, the time limit was set to 24 hours on Pentium 2.2GHz machines with 2GB memory. For the second procedure, the time limit was 90 minutes on Pentium 3.0GHz machines with 6GB memory. The MDS analysis was also applied.

\begin{center}
\textit{Program Availability}
\end{center}
\indent\indent The rnasim program was written in ANSI C++ and runs on Linux platforms. It is available for downloading at http://kim.bio.upenn.edu/software.

\newpage
\bibliographystyle{sysbio}
\bibliography{refs}             

\begin{sidewaystable}
	\caption{Parsimony score and Robinson-Foulds (RF) distance for benchmark and simulated datasets}
		\begin{tabular}{ccccccccccc}
	\hline
 	Gapped  & \multicolumn{5}{c}{Parsimony Score} & \multicolumn{5}{c}{RF Distance} \\
	\cline{2-11}
	Datasets 	 & maximum & minimum & range & average & S.D. & maximum & minimum & range & average  & S.D. \\
	\hline
	Benchmark & 78818 & 78662 & 156 & 52.4 & 21.8 & 533  & 186  & 347 & 369.2 & 36.9\\
	\hline
	rnasim & 56617 & 56534 & 83 & 17.9 & 9.4 & 232 & 42 & 190 & 130.6 & 21.0 \\
	\hline
	ROSE & 52776 & 52753 & 23 & 2.6 & 2.9 & 92 & 9 & 83 & 41.9 & 9.5 \\
	\hline	
	\hline
 	Ungapped & \multicolumn{5}{c}{} & \multicolumn{5}{c}{}\\
	Datasets & \multicolumn{5}{c}{} & \multicolumn{5}{c}{}\\
	\cline{2-11}			
	\hline
	Benchmark & 8992 & 8957 & 35 & 15.7 & 6.0 & 591 & 307 & 284 & 471.8 & 29.1\\
	\hline
	rnasim & 9285 & 9271 & 14 & 4.3 & 3.2 & 195 & 58 & 137 & 128.6 & 17.0 \\
	\hline
	ROSE & 8714 & 8700 & 14 & 1.9 & 1.9 & 155 & 37 & 118 & 85.5 & 13.2 \\
	\hline
	Seq-Gen & 9813 & 9799 & 14 & 2.2 & 1.8 & 125 & 30 & 95 & 44.6 & 11.4 \\
	\hline
	\end{tabular}

\end{sidewaystable}
\label{table1}

\begin{sidewaystable}
	\caption{Likelihood score and Robinson-Foulds (RF) distance for benchmark and simulated datasets}
		\begin{tabular}{ccccccccccc}
	\hline
 	Gapped  & \multicolumn{5}{c}{Likelihood Score} &  \multicolumn{5}{c}{RF Distance} \\
	\cline{2-11}
	Datasets 	 & maximum & minimum & range & average & S.D. & maximum & minimum & range & average  & S.D. \\
	\hline
	Benchmark	& 76119.50 & 68635.10 & 7484.40 & 1453.06 & 1392.00 & 158 & 0 & 158 & 75.49 & 18.88\\
	\hline
	rnasim	& 61343.65 & 55237.00 & 6106.65 & 1301.35 & 1138.61 & 134 & 0 & 134 & 73.16 & 16.04\\
	\hline
	ROSE	& 56004.08 & 54467.53 & 1536.55 & 14.26 & 63.93	& 48 & 0 & 48 & 6.22 & 10.48\\
	\hline	
	\hline
 	Ungapped & \multicolumn{5}{c}{} & \multicolumn{5}{c}{}\\
	Datasets & \multicolumn{5}{c}{} & \multicolumn{5}{c}{}\\
	\cline{2-11}			
	\hline
	Benchmark & 	16318.98 & 13309.44 & 3009.54 & 838.65	& 558.22 & 174 & 33 & 141 & 125.05 & 17.04 \\
	\hline
	rnasim	&	12337.76 & 10422.86 & 1914.90 & 311.60 & 412.06 & 143 & 0 & 143 & 77.35 & 22.89 \\
	\hline
	ROSE	&	10741.27 & 10741.27 & 0.00 & 0.00 & 0.00 & 0 & 0 & 0 & 0 & 0 \\
	\hline
	Seq-Gen	&	11859.41 & 11859.01 & 0.40 & 0.18 & 0.20 & 7 & 0 & 7 & 3 & 3 \\
	\hline
	\end{tabular}
\end{sidewaystable}
\label{table2}

\newpage
\section*{Figure Captions}

FIGURE1. The sequence and secondary structure of an \emph{E. coli} 5S rRNA. The parenthesis format is a convenient way to record the secondary structure. A left parenthesis and its matching right parenthesis denote a base pairing. \\

\noindent FIGURE 2. The RNA evolution simulation scheme. The molecule in the center of a circle denotes a wildtype RNA with its folding free energy in the adjacent parenthesis. All other molecules in the circle are mutants that can  mutate from the wildtype RNA by a single mutation that can be an insertion, a deletion, or a substitution. The mutant ensemble consists of all mutants in the circle. Each mutant has its own folding free energy. The change of folding free energy, $\Delta E$, determines a mutant's fitness change from the wildtype, therefore, its fixation rate $r$ (see text for details). A mutant with the lowest folding free energy has the maximal fixation rate $r_{max}$.  In the simulation, a mutant is randomly drawn from the mutant ensemble and is accepted with probability proportion to its fixation rate. For example, M$_1$ is the fixed mutant for M, and M$_2$ is for M$_1$ (For clarity, a circle representing M$_2$'s mutant ensemble is not drawn). Once a mutant is fixed, it acquires its own mutant ensemble, and the sampling procedure is conducted to selecte the next fixed mutant.\\

\noindent FIGURE 3. The folding free energy decrease ($\Delta E < 0$ in Equation 1) between a mutant and its wildtype RNA during a simulation on a 5000-taxon ultrametric binary tree by depth-first traversal. The horizontal axis records the sequentially observed mutants (may not fix)  with negative $\Delta E$. The minimum of $\Delta E$ is -8.2\emph{J/mol}. \\

\noindent FIGURE 4. Distributions of $\Delta E$, the folding free energy change  between \emph{E.coli} 16S rRNA and its mutants in the mutant ensemble $\mathscr{E}$ (see text for definition). The substitution mutants differ from the wildtype by one nucleotide. The deletion and insertion mutants are converted from the wildtype RNA by deleting or inserting a stretch of 1 to 7 nucleotides. The ranges of $\Delta E$ for the three mutation types are $-6.6$ to $7.9$ for substitution, $-9.6$ to $ 12.4$ for deletion and  $-7.2$ to  $8.0$ for insertion. If the indel size is restricted to one, the ranges are  $-9.6$ to $9.7$ for deletion and $-7.1$ to $5.8$ for insertion. $\Delta E$ is measured in \emph{J/mol}.\\

\noindent FIGURE 5. Fixation rate as a function of fitness change and population size (Equation 4). The upper half corresponds to fitness increase ($s>0$),  the bottom half to fitness decrease ($s<0$), and the middle line to no fitness change ($s=0$). The graph is drawn after Figure 7-2 in \citep[page 187]{crow1986bcp}.\\

\noindent FIGURE 6. The effect of population size on RNA evolution. Left: the \emph{E. coli} small-subunit rRNA used as root molecule in simulations. Middle: a typical simulated RNA when population size $N=10000$.  Right: a typical simulated RNA when population size $N=10$, the base pairings are basically decoupled.\\

\noindent FIGURE 7. In PAUP*'s heuristic search by TBR branch swapping, different datasets take different time to decrease by 10\%, 50\%, 95\%, and 99\% of the total parsimony score decrement, i.e., the difference between the starting parsimony score and the final parsimony score. The abscissa coordinates use logarithmic values with base 2. (a) gapped sequences for benchmark, rnasim, and ROSE datasets with average pairwise identity $ 73.0\pm1.0$\% and minimal pairwise identity $ 47.0\pm1.0$\%.  (b) ungapped sequences for benchmark, rnasim, ROSE and Seq-Gen datasets with average pairwise identity $87.0\pm 1.0$\% and minimal pairwise identity $ 62.0\pm1.0$\%. (c) ungapped sequences for rnasim, ROSE and Seq-Gen datasets with average pairwise identity $93.0\pm 1.0$\% and minimal pairwise identity $ 75.0\pm1.0$\%. For both gapped and ungapped comparisons, there is one empirical benchmark dataset and three datasets for each simulator. Each dataset was tree searched 10 times. See Materials and Methods for details on sequence generation and evaluation.\\

\noindent FIGURE 8.  The relative distance of the 1000 trees found in PAUP*'s heuristic search for ungapped datasets with parsimony criterion (a: benchmark, b: rnasim, c: ROSE, d: Seq-Gen) and likelihood criterion (e: benchmark, f: rnasim, g: ROSE, h: Seq-Gen) . Each graph plots the first two dimensions of a multidimensional scaling on the RF distance matrix of a dataset. Graph a-d used the empirical dataset and one dataset for each simulator in generating Figure~\ref{Figure7}b, while Graph e-f used only 100 sequences from a corresponding dataset.\\

\noindent FIGURE 9. In PAUP*'s heuristic search by TBR branch swapping, different datasets take different time to decrease by 10\%, 50\%, 95\%, and 99\% of the total likelihood score decrement, i.e., the difference between the starting likelihood score and the final likelihood score.  The abscissa coordinates use logarithmic values with base 2. (a) gapped sequences for benchmark, rnasim, and ROSE datasets. Each dataset consists of 100 sequences sampled from the datasets used in generating Figure~\ref{Figure7}a. (b) ungapped sequences for benchmark, rnasim, ROSE and Seq-Gen datasets. Each dataset consists of 100 sequences sampled from the datasets used in generating Figure~\ref{Figure7}b. For both gapped and ungapped comparisons, there is one empirical benchmark dataset and three datasets for each simulator. Each dataset was tree searched 10 times.

\newpage

\newpage
 .
\newpage

\newpage
 .
\newpage

\newpage
 .
\newpage

\newpage
 .
\newpage

\newpage
.
\newpage
\begin{figure}
\begin{center}
\epsfig{file=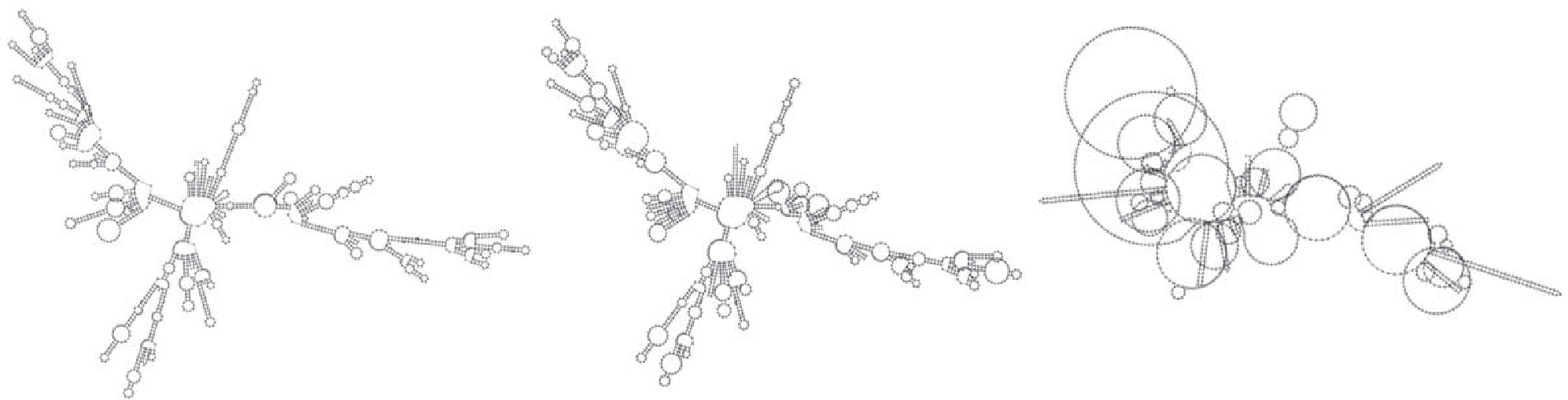, width=1.0\linewidth, angle=0}
\end{center}
\caption{\label{Figure6}}
\end{figure}

\newpage
.
\newpage
\begin{figure}
\centering
\begin{tabular}{ccc}
\epsfig{file=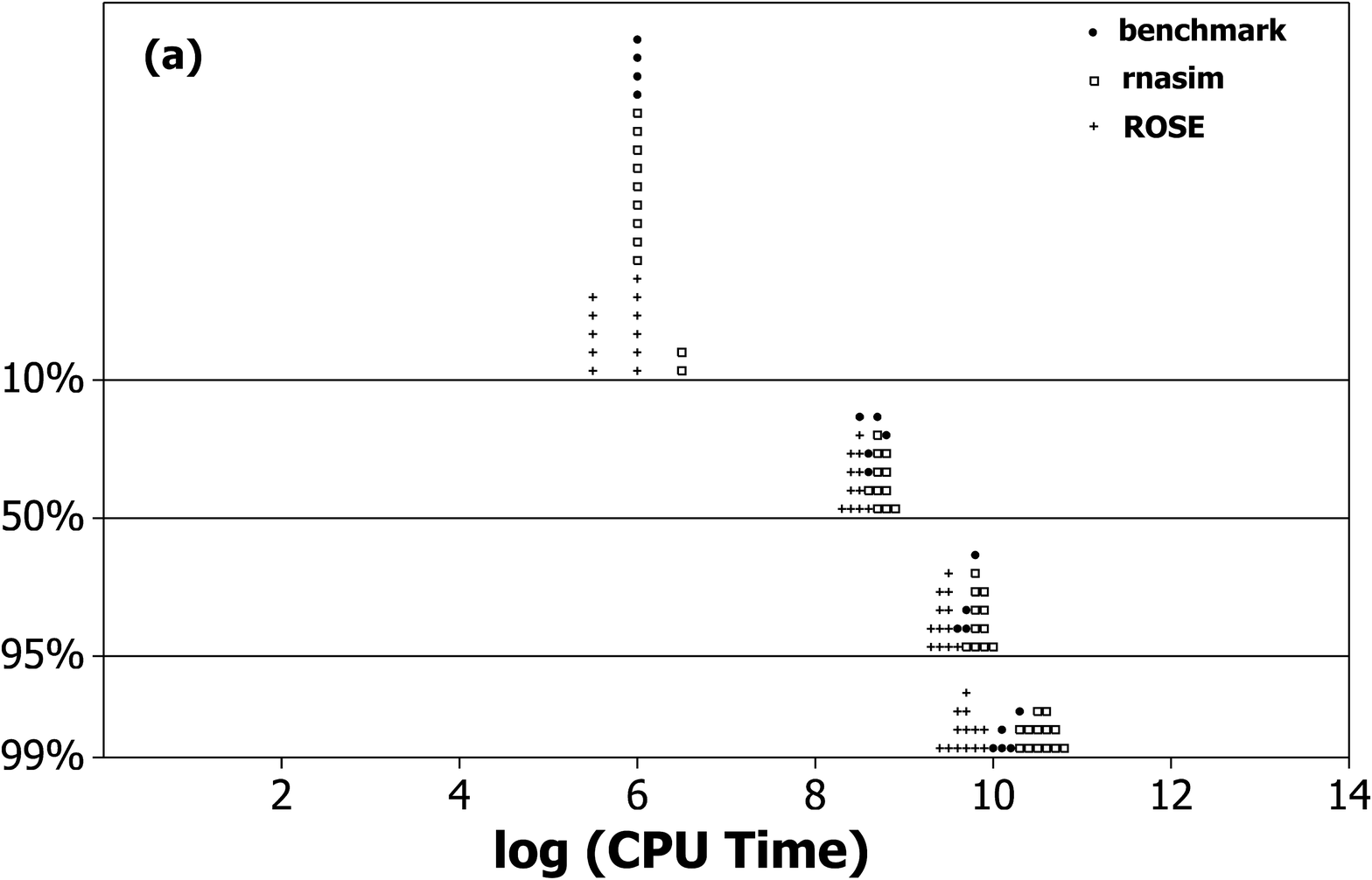,width=0.8\linewidth, angle=0} \\
\epsfig{file=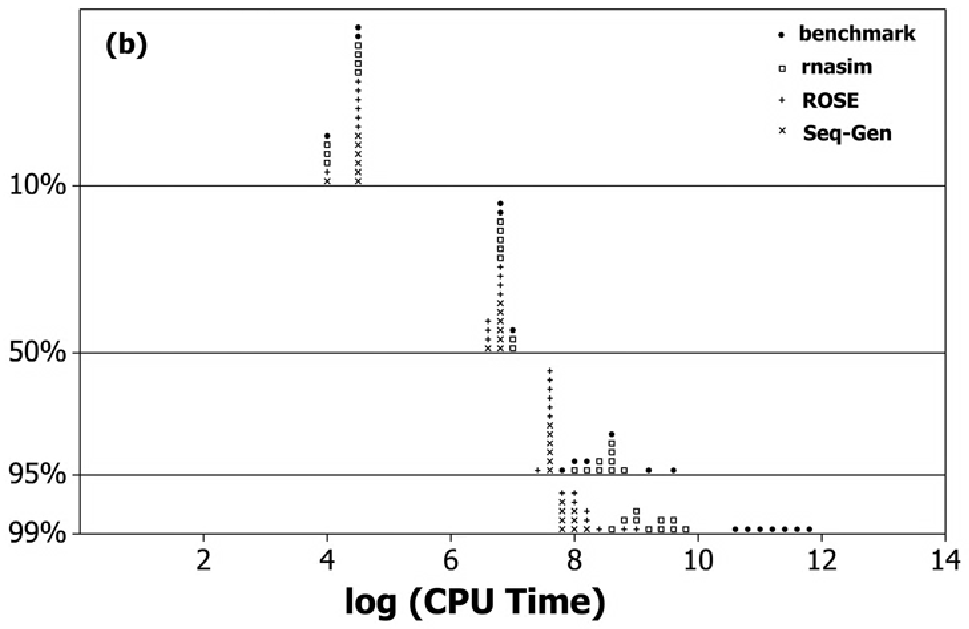,width=0.8\linewidth, angle=0} \\
\epsfig{file=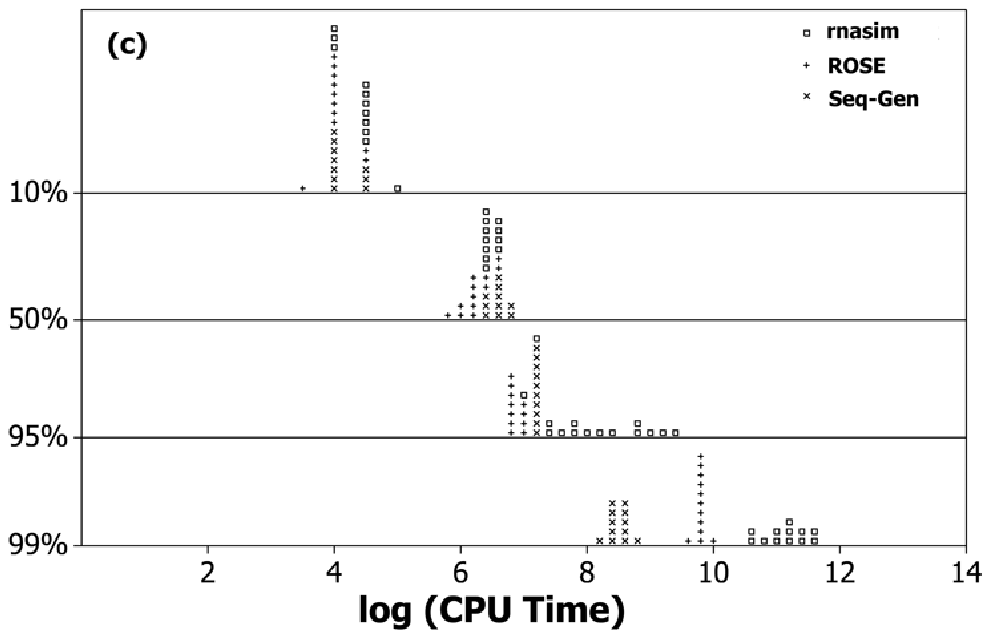,width=0.8\linewidth, angle=0}
\end{tabular}
\caption{\label{Figure7}}
\end{figure}

\newpage
\begin{figure}
\centering
\begin{tabular}{cccc}
\epsfig{file=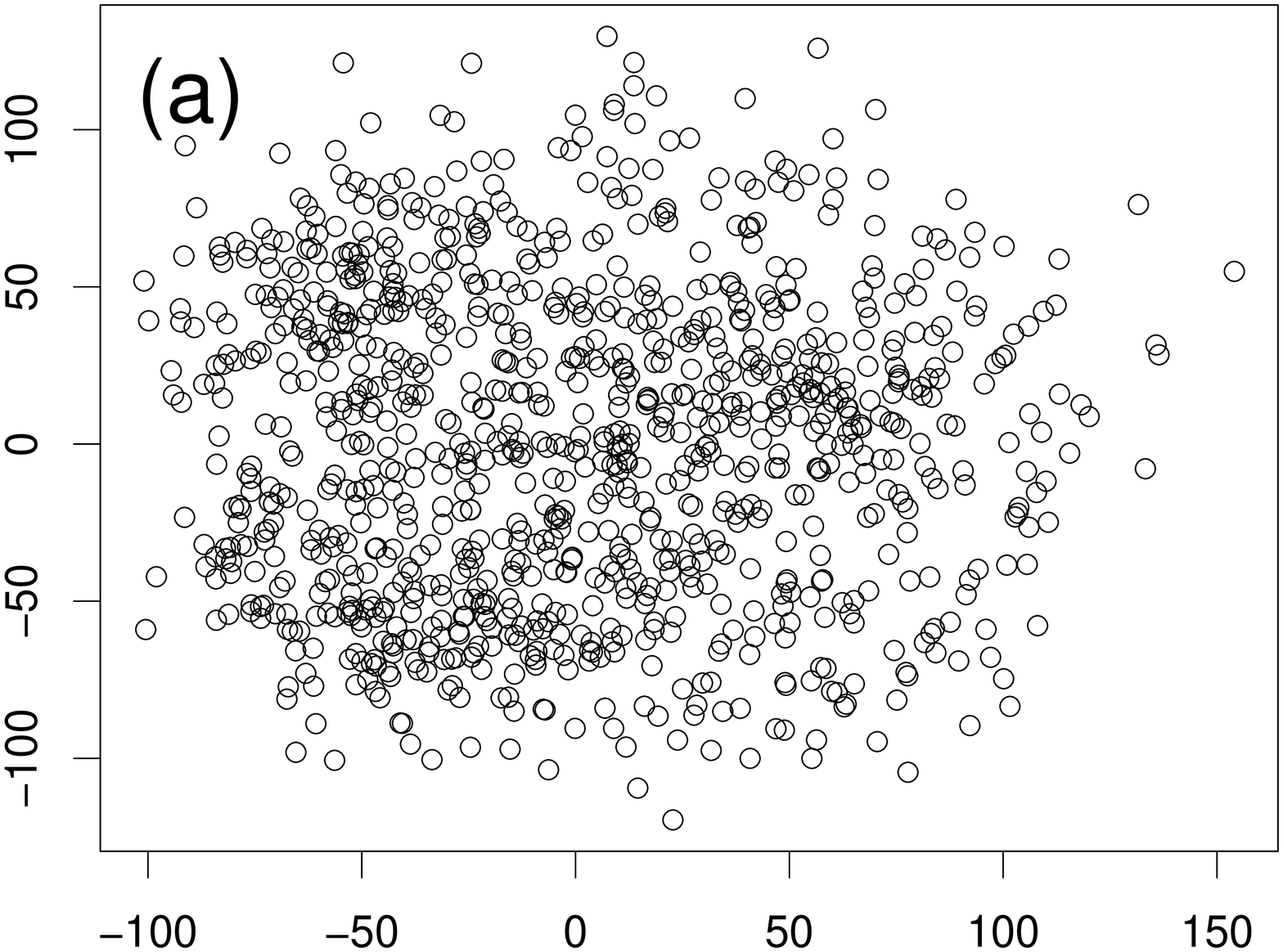,width=0.3\linewidth, angle=-90}
\epsfig{file=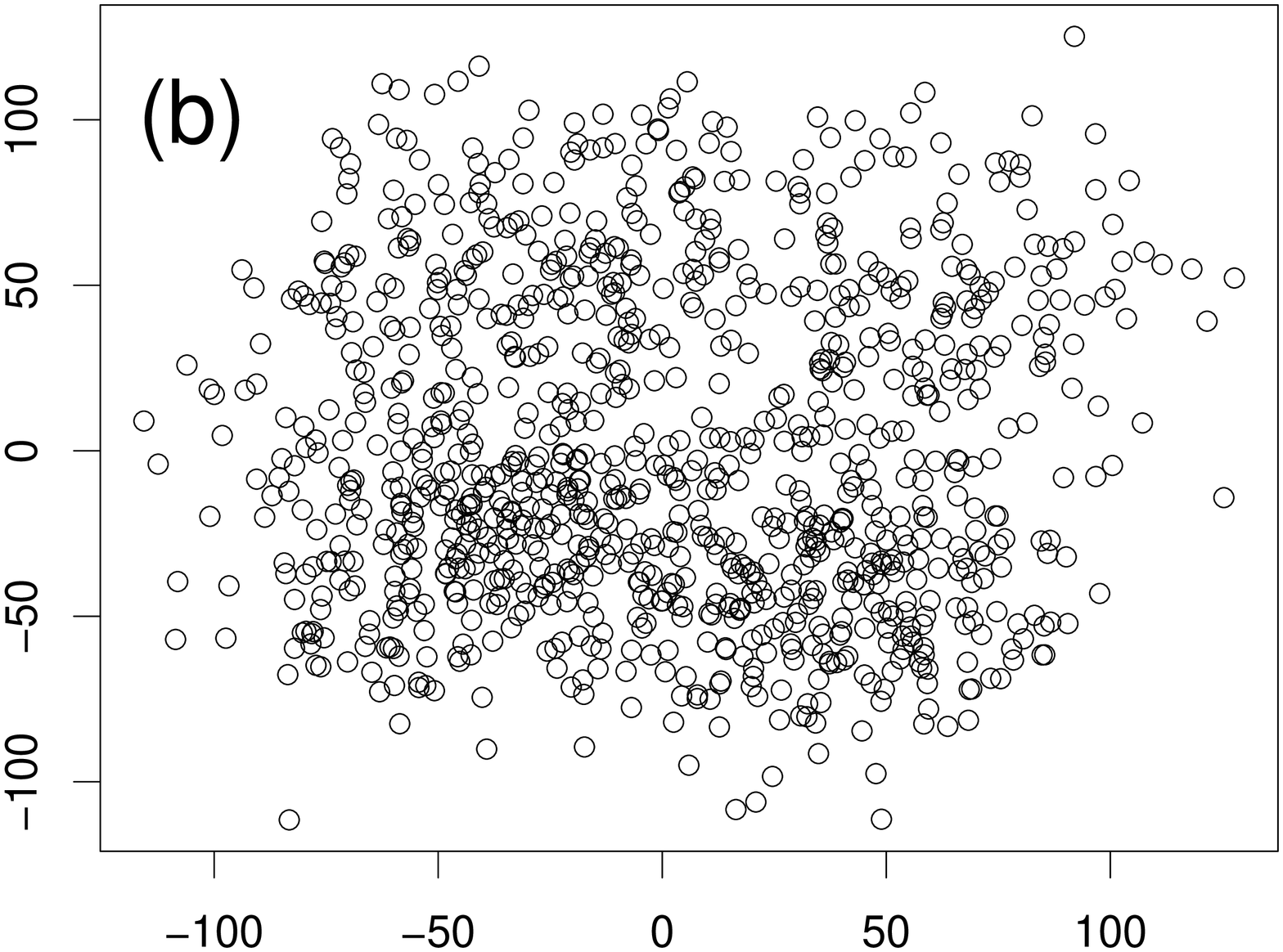,width=0.3\linewidth, angle=-90} \\
\epsfig{file=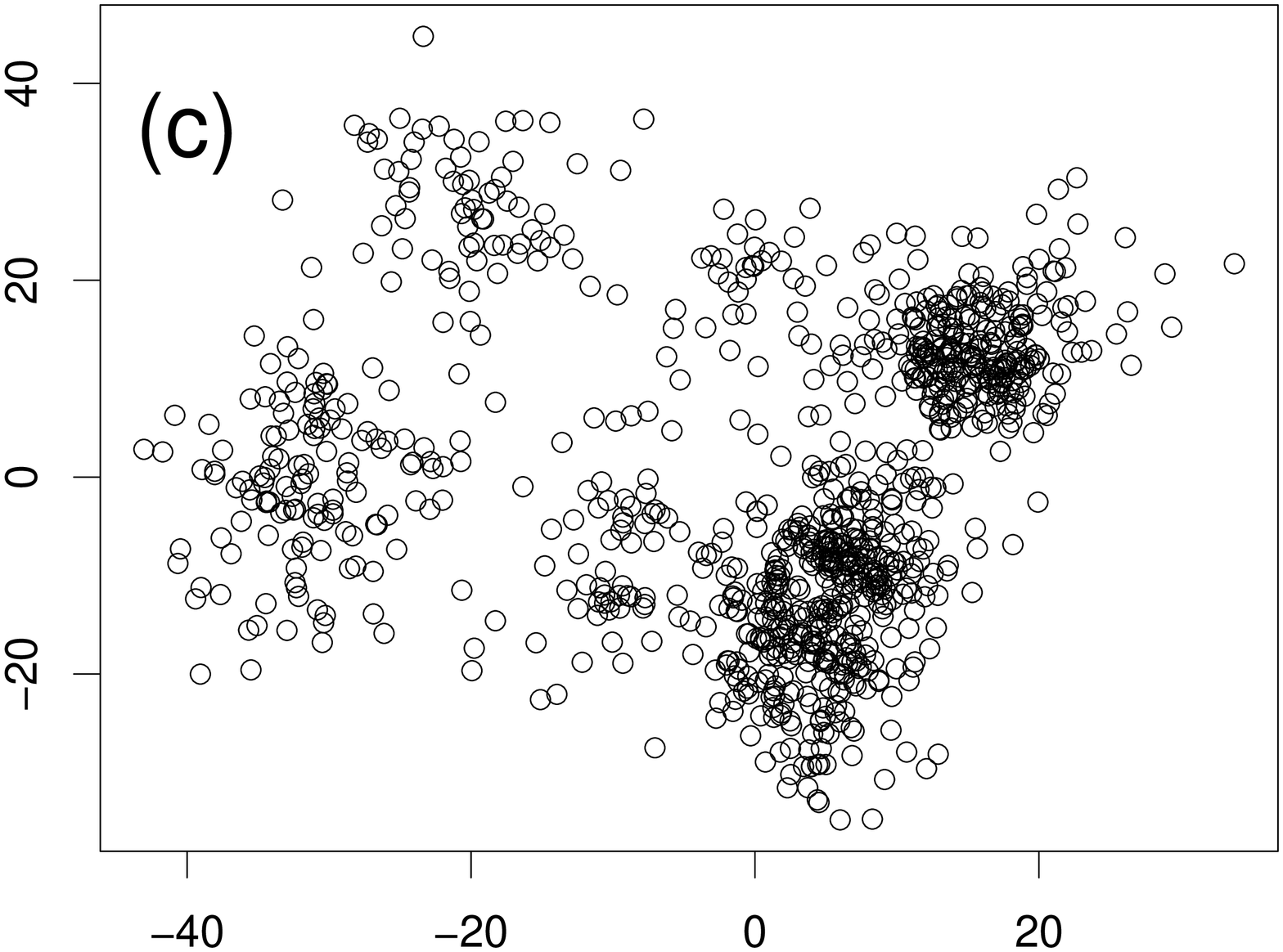,width=0.3\linewidth, angle=-90}
\epsfig{file=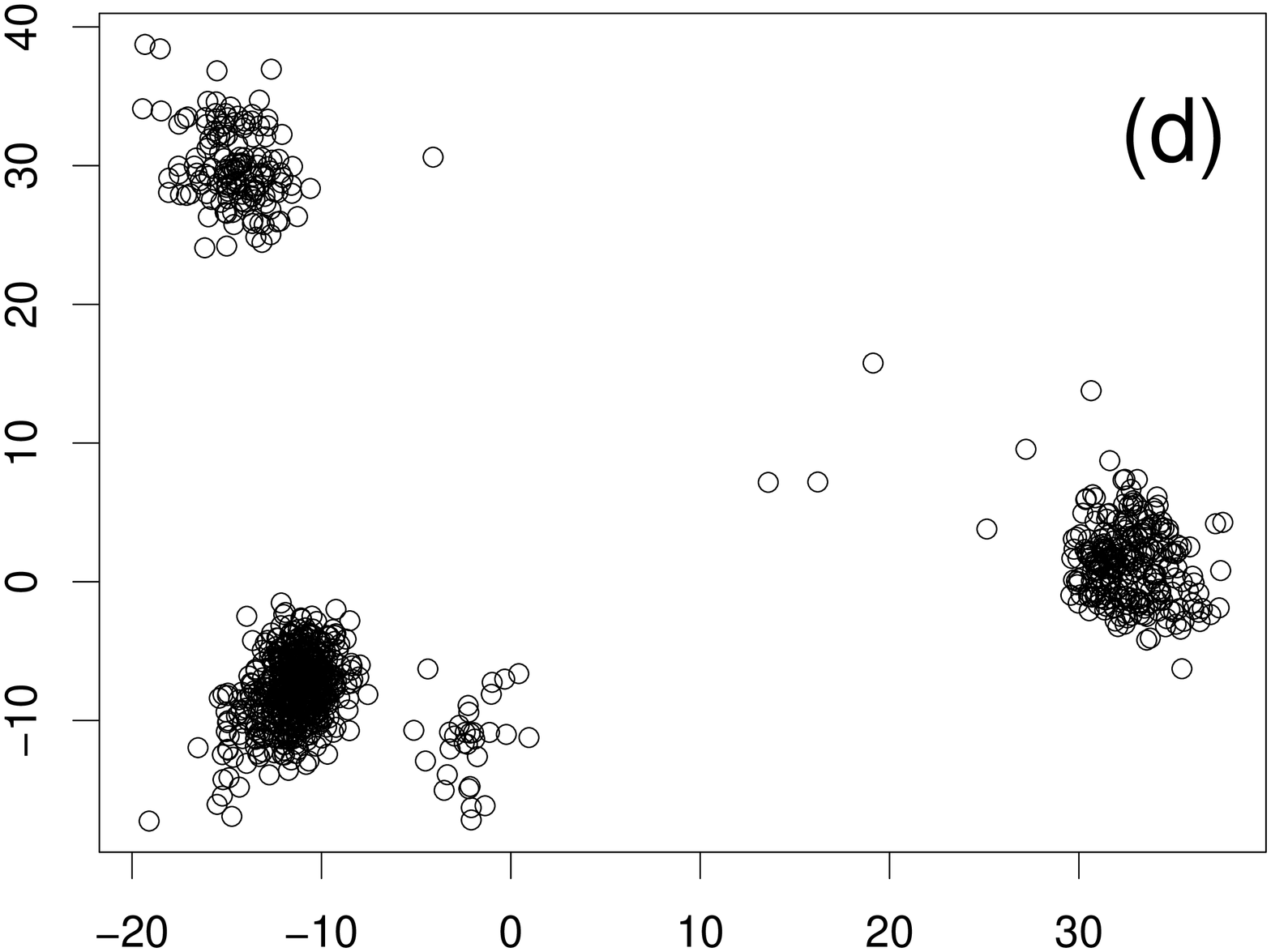,width=0.3\linewidth, angle=-90} \\
\epsfig{file=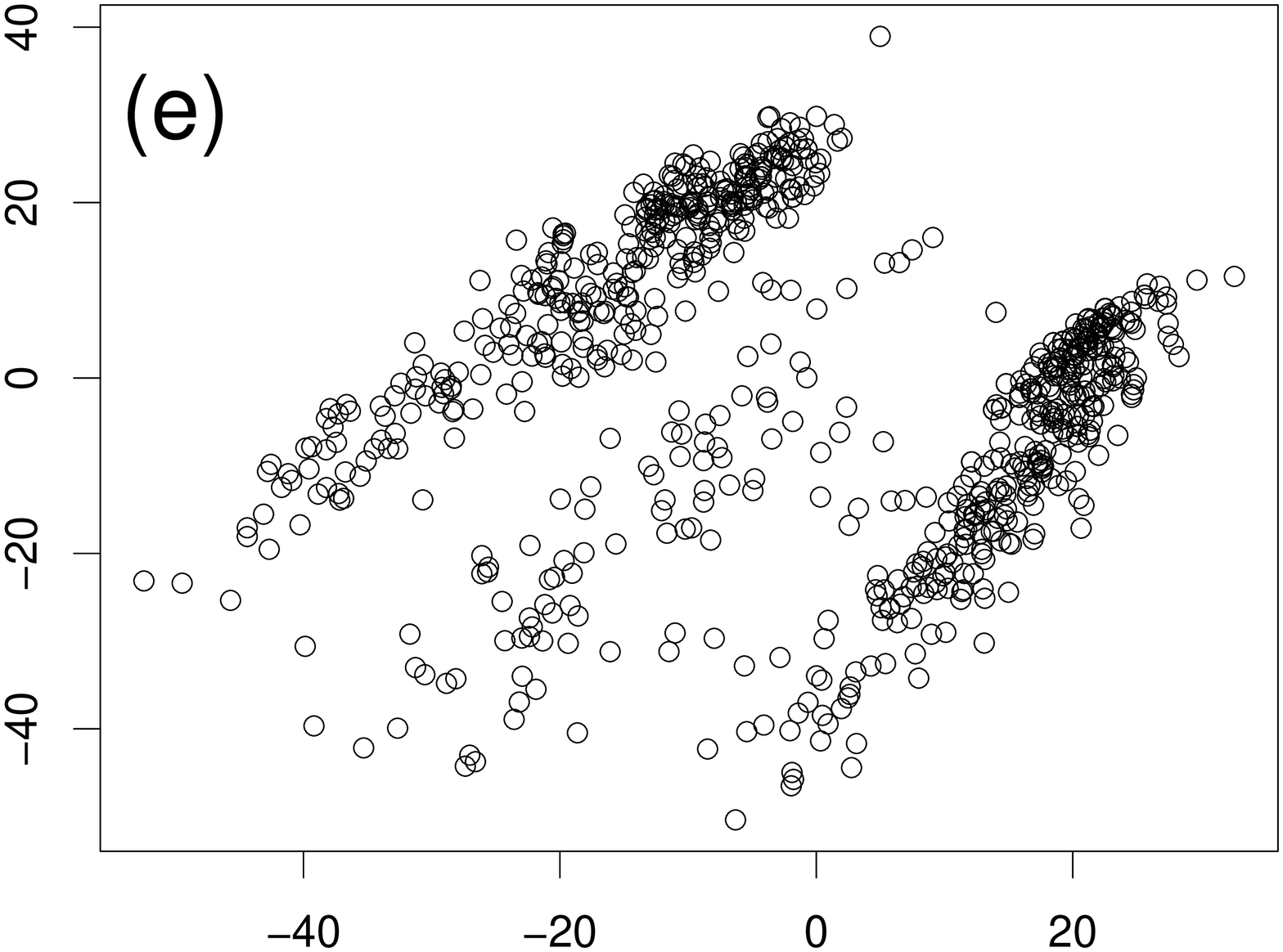,width=0.3\linewidth, angle=-90}
\epsfig{file=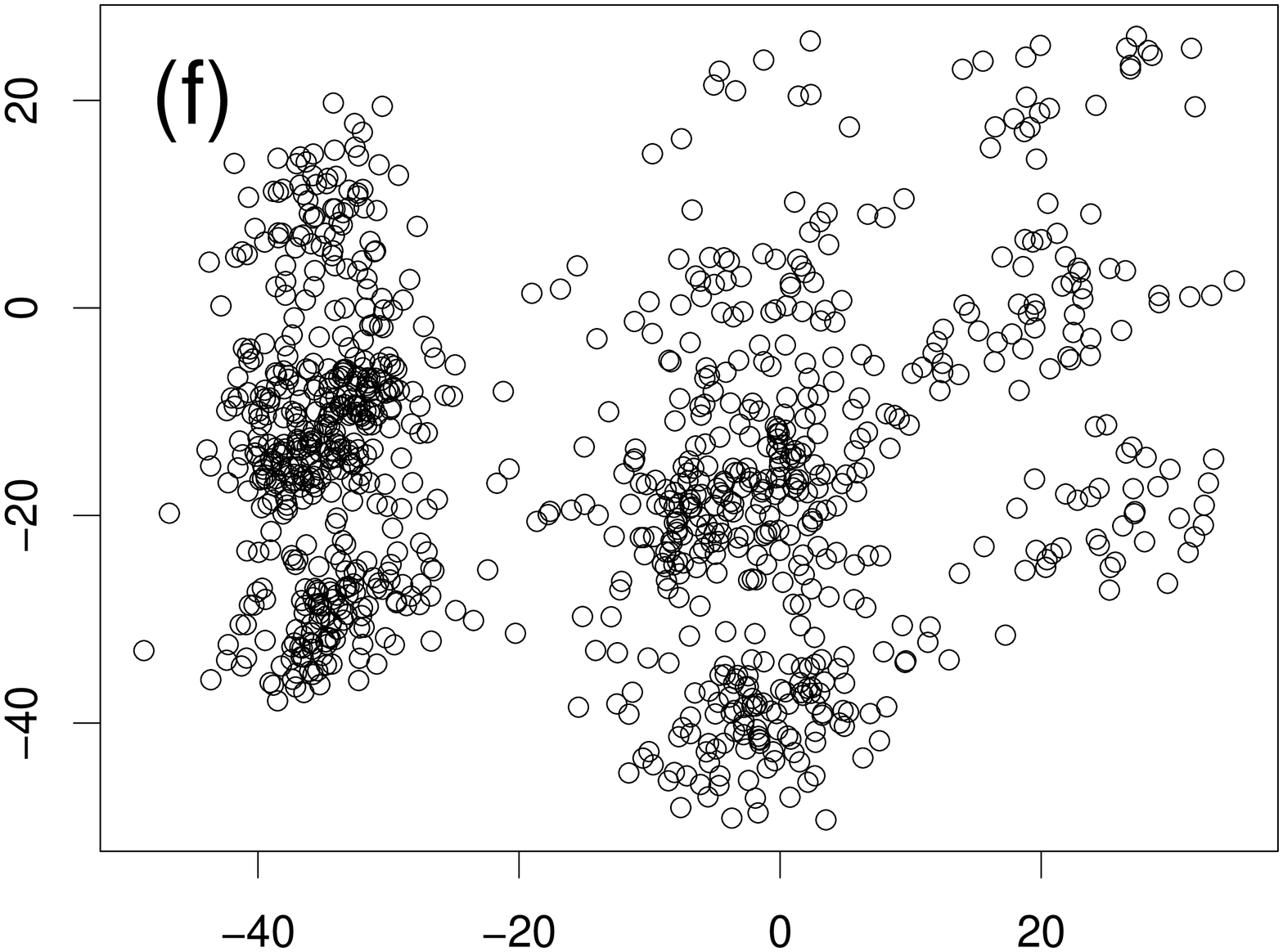,width=0.3\linewidth, angle=-90} \\
\epsfig{file=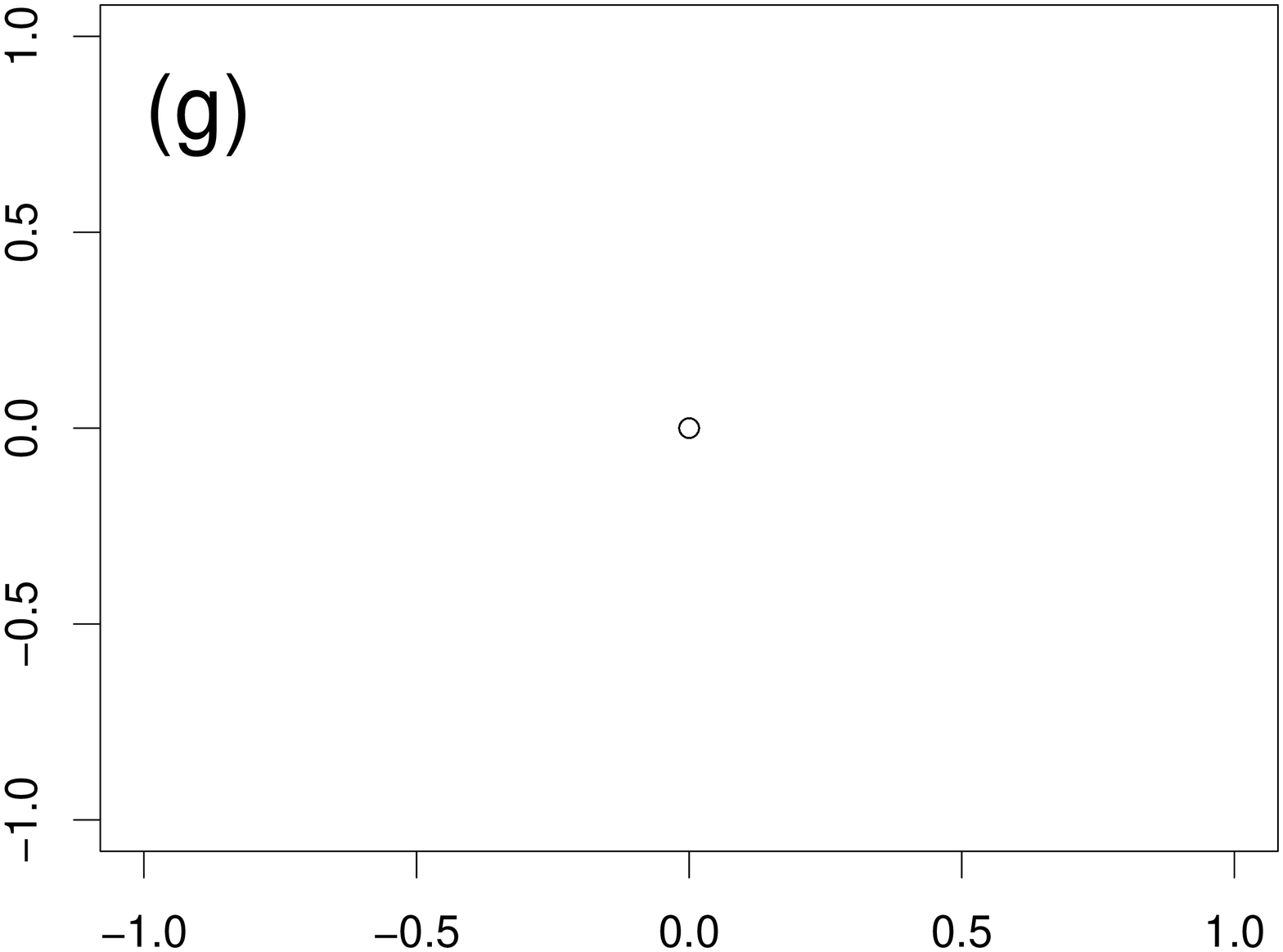,width=0.3\linewidth, angle=-90}
\epsfig{file=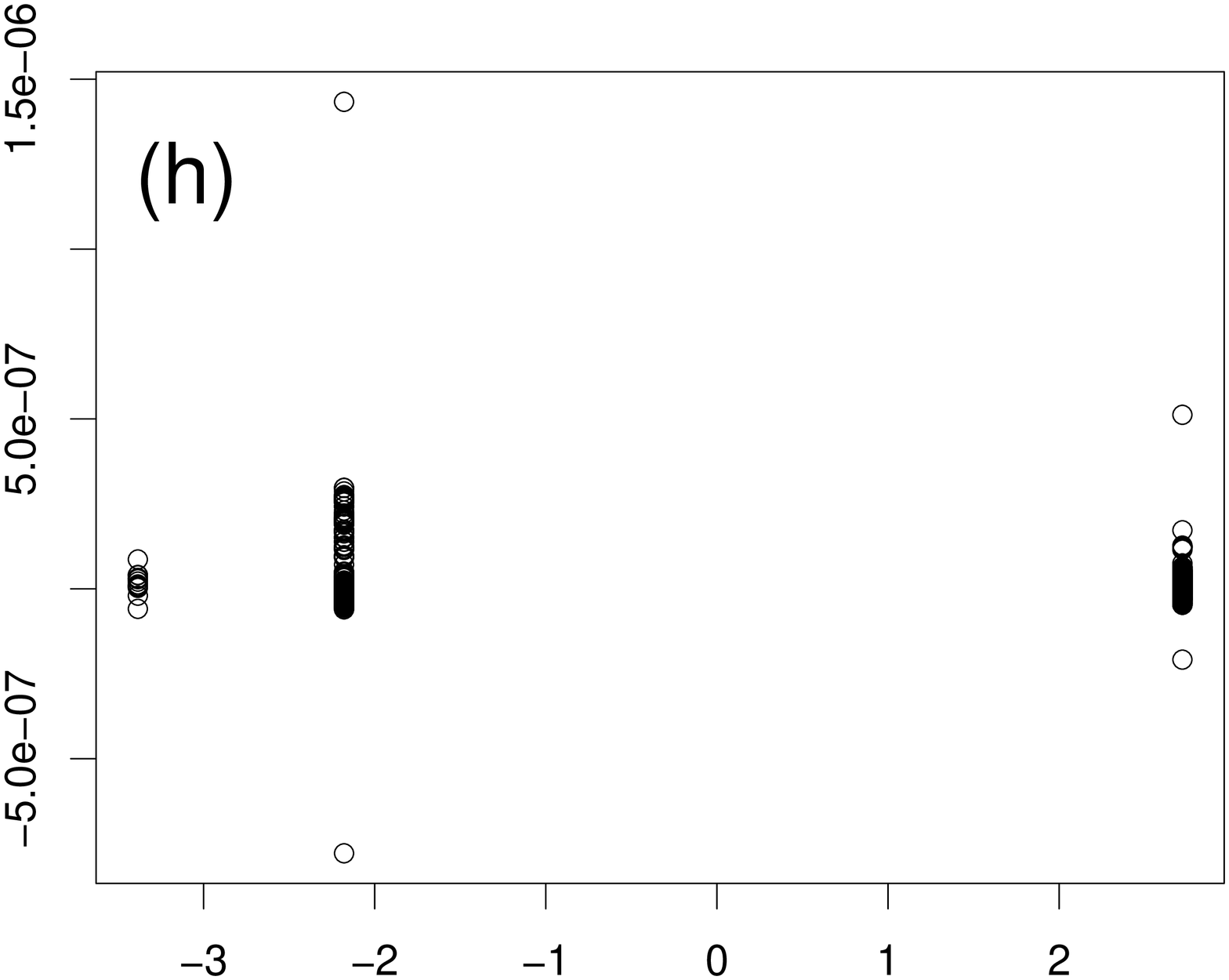,width=0.3\linewidth, angle=-90}
\end{tabular}
\caption{\label{Figure8}}
\end{figure}

\newpage
\begin{figure}
\centering
\begin{tabular}{cc}
\epsfig{file=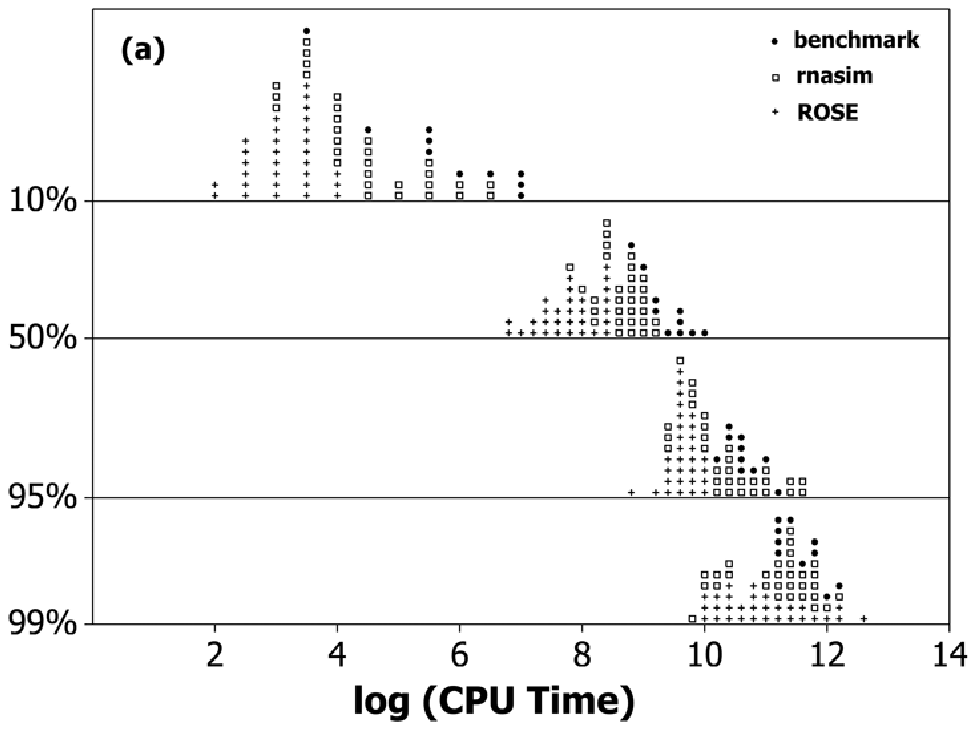,width=0.8\linewidth}\\
\epsfig{file=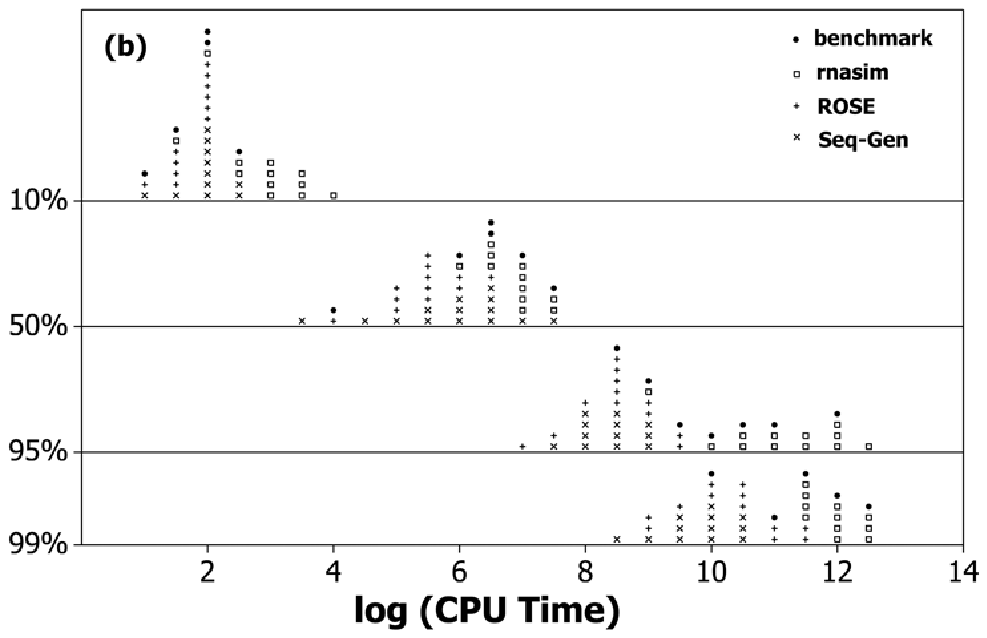,width=0.8\linewidth}
\end{tabular}
\caption{\label{Figure9}}
\end{figure}

\end{document}